\documentclass[a4paper,twocolumn,preprintnumbers,amsmath,amssymb,amsfonts,floatfix,nofootinbib,showkeys]{revtex4}
\usepackage{dcolumn}
\usepackage{bm}
\usepackage{float}
\usepackage{epsfig}
\usepackage{graphicx}
\usepackage{longtable} 
\usepackage{rotating}
\usepackage{amsmath}
\usepackage{amsfonts}
\usepackage{amssymb}
\usepackage{xspace}
\usepackage{multirow}
\usepackage{mathrsfs}
\usepackage{calrsfs}
\usepackage{txfonts}
\usepackage{graphicx}
\usepackage{hypernat}
\usepackage{color}
\usepackage[utf8]{inputenc}
\newif\ifcomment
\newif\ifprint
\graphicspath{{./pics/final/}}
\ifprint
 \usepackage[linkbordercolor={0 0 .8},%
             citebordercolor={.8 0 0},%
             urlbordercolor={.8 0 .8},%
             raiselinks=true,%
             pdfborder={0 0 1 [3]}]{hyperref}
\else
 \usepackage[pdftex,colorlinks, 
             linkcolor=blue,citecolor=blue,anchorcolor=blue,
             urlcolor=red]{hyperref}
\fi
\pdfoutput=1        
\pdfcatalog{/PageMode(/UseOutlines)} 
\newcommand{\pp}           {pp}

\newcommand{\pA}           {p-A}
\newcommand{\dAu}          {d-Au}

\newcommand{\AuAu}         {Au-Au}
\newcommand{\SuSu}         {S-S}
\newcommand{\SAu}          {S-Au}

\newcommand{\pT}           {\ensuremath{p_{\rm T}}}

\newcommand{\sqrts}        {\ensuremath{\sqrt{s}}}

\newcommand{\hrefurl}[1]   {\href{#1}{\url{#1}}}
\newcommand{\Ref}[1]       {Ref.~\cite{#1}}

\newcommand{\Tab}[1]       {Tab.~\ref{#1}}
\newcommand{\Fig}[1]       {Fig.~\ref{#1}}
\newcommand{\Figs}[2]      {Figs.~\ref{#1} and \ref{#2}}
\newcommand{\Figsm}[2]     {Figs.~\ref{#1}--\ref{#2}}

\newcommand{\Eq}[1]        {Eq.~\ref{#1}}
\newcommand{\Sec}[1]       {Sec.~\ref{#1}}

\newcommand{\co}[1]        {}
\newcommand{\ignore}[1]    {}
\newcommand{\pt}           {p_{\text{T}}}
\newcommand{\mt}           {m_{\text{T}}}

\begin{document}
\title{Applicability of transverse mass scaling in hadronic collisions at the LHC}
\date{\today}
\author{Lucas Altenk\"amper}\affiliation{UiB, Bergen, Norway}
\author{Friederike Bock}\affiliation{CERN, Geneva, Switzerland}
\author{Constantin Loizides}\affiliation{ORNL, Oak Ridge, USA}
\author{Nicolas Schmidt}\affiliation{ORNL, Oak Ridge, USA\\}
\begin{abstract}\noindent
We present a study on the applicability of transverse mass scaling for identified particle spectra in proton-proton collisions at $\sqrts=7$~TeV based on data taken by the ALICE experiment at the LHC. 
The measured yields are parametrized and compared to estimates obtained from a generalized transverse mass scaling approach applied to different reference particle spectra.
It is found that generalized transverse mass scaling is not able to describe the measured spectra over the full range in transverse momentum. 
At low $\pt$, deviations of $20$\% or more are obtained, in particular if pions are used as reference particles. 
A better scaling performance is obtained, when kaons are used as reference particles.
At high $\pt$ all tested spectra with the possible exception of the charged kaons exhibit a scaling behavior.
Investigating the feed-down contributions from resonance decays to the charged pion yields reveals, that using them as reference a general scaling may not be achievable.
Our findings imply that for precision measurements of direct photon and di-electron spectra at low transverse momentum one should measure the relevant hadronic background, instead of relying on $\mt$ scaling for its estimate.
\end{abstract}
\maketitle
\section{Introduction}
Already as early as 1965, Hagedorn proposed that the transverse mass spectra of hadrons in \pp\ collisions should be governed by a universal scaling law, based on a statistical-thermodynamical approach for hadron production~\cite{Hagedorn:1965st}. 
The spectra, expressed as a function of the transverse mass $\mt = \sqrt{\pt^2 + m_0^2}$ ~(where $\pt$ is the transverse momentum and $m_0$ the rest mass), should follow an exponential distribution with a universal inverse slope parameter. 
This parameter was originally identified as the highest possible temperature for hadronic matter.
Indeed, it was found that the $\mt$ spectra of different particle species (e.g.\ $\pi^-$, $\overline{\text{K}}^0$, $\Lambda$, $\rho^0$ and $\omega$) produced in hadronic collisions at center-of-mass energies of $8$--$16$ GeV are reasonably well described by an exponential or Bose-Einstein distribution with the inverse slope parameter being independent of the particle mass~\cite{Deutschmann:1974ne, Bartke:1976zj}. 
Subsequently, the effect was called ``$\mt$-scaling''~\cite{Gatoff:1992cv}.

Furthermore, it was also shown that the data from \pp\ collisions over a range of about $6$ to $63$~GeV can be parametrized with a generalized function of the transverse mass~\cite{Alper:1974rw,Bourquin:1976fe}.
The generalized function, however, assumes a factorization of the rapidity~($y$) and $\mt$ dependence and is in contrast to the statistical-thermodynamical approach not of purely exponential form.
A phenomenological scaling law was also observed at the CERN SPS by the WA80 collaboration in \SuSu\ and \SAu\ collisions at center-of-mass energies of $200$~GeV per nucleon pair~\cite{Albrecht:1995ug}.
The transverse mass spectra of the $\eta$ and $\pi^0$ mesons were found to be identical up to a constant factor in both systems. 
However, in contrast to the production at lower collision energies the spectra were best described by a power law, which differs from the universal scaling originally proposed by Hagedorn.
The approximate $\mt$-scaling was further tested for mesons and baryons by the STAR collaboration in \pp\ collisions at $\sqrt{s}=200$~GeV~\cite{Abelev:2006cs}, and a difference between mesons and baryons was observed for the first time with the meson spectra appearing to be systematically harder than the baryon spectra for transverse masses above about $2$~GeV. 
The difference between the exact $\mt$-scaling, as originally proposed by Hagedorn, and the more general but approximate form was also investigated by the PHENIX collaboration in \pp\ collisions at $\sqrt{s}=200$~GeV~\cite{Adare:2010fe}.
The spectra were found to obey the scaling relation separately for mesons and baryons, but in accordance with \cite{Abelev:2006cs} no scaling relation in between mesons and baryons was observed. 
Additional tests based on RHIC data were performed in \Ref{Khandai:2012xx} for different collision systems and energies using a modified Hagedorn parameterization~(which is close to an exponential at low $\pt$ and an exact power law at high $\pt$).
It was found that the $\mt$ spectra of mesons ($\phi$, K$^{\pm}$ and K$^0_S$) and baryons ($\Lambda$, $\Delta$, $\Xi$ and $\Omega$) could be described separately by applying the approximate transverse mass scaling relation of the pion or proton, respectively, in \pp\ and \dAu\ collisions at $\sqrt{s}=200$~GeV. 
However, a disagreement was observed\co{ for the $\Lambda$ baryon at high $\pt$ in \AuAu\ collisions at $\sqrt{s}=200$~GeV and} for $\phi$ mesons and kaons in \pp\ collisions at $\sqrt{s}=900$~GeV, indicating that the generalized scaling breaks down at higher collision energies.

Originally, the phenomenological scaling law was observed directly on $\mt$ differential yields. 
However, it is more practical to formulate the scaling as a function of $\pt$, by rewriting the functional form of the parametrization, that is used as a basis for the scaling, in terms of $\mt$ rather than $\pt$ through the substitution
\begin{align}
	\pt \rightarrow \sqrt{\mt^2 - m_0^2}\,.
	\label{eq:substitution}
\end{align}
In this way, since the corresponding invariant yields are equal, 
\begin{align}
	\frac{1}{\pt} \frac{\text{d}^2N}{\text{d}\pt\text{d}y} = \frac{1}{\mt} \frac{\text{d}^2N}{\text{d}\mt\text{d}y}\,,
	\label{eq:approxEqual}
\end{align}
the scaling can be expressed in terms of transverse momentum where it must be ensured that both spectra\co{, i.e.\ the basis and the scaled spectrum,} are evaluated at the same transverse mass. 
Hence, at midrapidity, the invariant $\pt$ spectrum of a particle of type $X$, $f^{\text{inv.}}_{\rm X}(\pt)$, can be obtained by scaling the parametrization of the invariant transverse momentum spectrum of another particle $R$, $f^{\text{inv.}}_{\rm R}(\pt)$, used as the reference, by
\begin{align}
	f^{\text{inv.}}_{\rm X}(p_{\text{T,X}}) &= C^{\rm X}_{\rm R} \times f^{\text{inv.}}_{\rm R}\left(\sqrt{m_{\text{T,X}}^2 - m_{0,{\rm R}}^2}\right)  \nonumber \\
								&= C^{\rm X}_{\rm R} \times f^{\text{inv.}}_{\rm R}\left(\sqrt{m_{0,{\rm X}}^2 + p_{\text{T,X}}^2 - m_{0,{\rm R}}^2}\right),
	\label{eq:scalingRelationInvariant}
\end{align} 
where the substitution from \Eq{eq:substitution} and the evaluation of both spectra at the same transverse mass,
\begin{align} 
	p_{\text{T,X}}^2 + m_{0,{\rm X}}^2 = p_{\text{T,R}}^2 + m_{0,{\rm R}}^2,
\end{align}
were applied to the initial (reference) spectrum. 
The parameter $C^{\rm X}_{\rm R}$ denotes a constant offset between the spectra of the two particle species.
It can in principle be obtained from the data if the two particle spectra exhibit a similar power law at high $\pt$.
In the following we will drop the explicit notation $C_{\rm R}^{\rm X}$ and denote the constant scaling parameter simply with $C$.

The scaling relation for invariant, $\pt$-differential spectra can be translated into a relation for non-invariant spectra by multiplying both sides by their respective $\pt$,
\begin{align} 
	f_{\rm X}(p_{\text{T,X}}) = C &\times \frac{p_{\text{T,X}}}{\sqrt{m_{0,{\rm X}}^2 + p_{\text{T,X}}^2 - m_{0,{\rm R}}^2}} \nonumber\\
			     &\times f_{\rm R}\left(\sqrt{m_{0,{\rm X}}^2 + p_{\text{T,X}}^2 - m_{0,{\rm R}}^2}\right),
	\label{eq:scalingRelation}
\end{align}
where the notation ``inv.'' was omitted to denote the change to non-invariant spectra.
Formulating approximate $\mt$-scaling in this way is operationally similar to the generalized approach discussed in \Ref{SchaffnerBielich:2001qj}. 

Collective effects, such as radial flow known to be present in heavy-ion, but also in \pA\ and perhaps even in \pp\ collisions~\cite{Loizides:2016tew}, should modify the scaling relation to some degree.
This can be demonstrated by assuming a simple model of flow, 
\begin{align}
	\pt \rightarrow \pt' = \pt + \beta m_0,
	\label{eq:radialFlow}
\end{align}
where the particles initially produced  with a transverse momentum $\pt$ are exposed to a boost by the radially expanding system with flow velocity $\beta$. 
Such a modification would introduce an additional mass dependence that affects different species to a different degree, in particular at low $\pt$. 
At sufficiently high $\pt$ any influence from such effects would however diminish.

Although transverse mass scaling can not be expected to hold in general, it is often applied to $\pt$ spectra in situations, where parametrizations of particle yields are required for which no measurements are available.
Estimates of hadronic backgrounds in direct photon~\cite{Adare:2012yt,Adare:2012vn,Afanasiev:2012dg,Adam:2015lda} or di-electron measurements~\cite{Adare:2009qk,Adamczyk:2015lme} are prominent examples where transverse mass scaling is employed. 
Furthermore, in such measurements, the parametrizations are usually extrapolated to lower and higher $\pt$.

In the following we will test the generalized $\mt$-scaling law on the basis of $\pt$ differential yields of identified hadrons measured by the ALICE experiment in \pp\ collisions at $\sqrt{s}=7$~TeV. 
For the purpose of this article, we focus on mesons, but the same study could be done for baryons as several different species were measured.
At first, we present in \Sec{sec:data} a large part of the available data and their parameterizations, followed by a discussion of the resonance feed-down contribution to the charged pion spectrum. 
Then in \Sec{sec:results} the measured yields are compared to the predictions obtained from applying the generalized $\mt$-scaling relation to the parametrizations of different spectra. 
Furthermore, they are compared to the particle ratios using the light mesons as reference particles.
We conclude in \Sec{sec:summary} with a summary of our findings.

\section{Available data and feed-down correction}
\label{sec:data}
\begin{figure*}
\begin{minipage}{\textwidth}
\begin{table}[H]
    \begin{tabular}{l c | c | c c c c c c c | l}
        \hline
        Hagedorn (\Eq{eq:modifiedHagedorn}) & Meas.\ range ($\text{GeV}/c$) & Fit range ($\text{GeV}/c$)& $A$ & $a$ & $b$ & $p_0$ & $n$ & & &Ref. \\ 
        \hline
        $\pi^0$                                                 & $0.3 < \pt < 25.0$& $0.3 < \pt < 16.0$ & $85.9\ignore{\pm 8.9}$ &$0.48 \ignore{\pm 0.02}$ &$-0.008 \ignore{\pm 0.02}$ &$0.59 \ignore{\pm 0.01}$ &$6.10 \ignore{\pm 0.03}$ & & & \cite{Abelev:2012cn} \\
        $\pi^{\pm}$                                             & $0.1 < \pt < 20.0$& $0.2 < \pt < 20.0$ &$98.1 \ignore{\pm 0.2}$ &$0.377 \ignore{\pm 0.004}$ &$-0.074 \ignore{\pm 0.002}$ & $0.556 \ignore{\pm 0.003}$&$6.04 \ignore{\pm 0.01}$ & & & \cite{Adam:2016dau} \\
        $\text{K}^{\pm}$                                        & $0.2 < \pt < 20.0$& $0.4 < \pt < 20.0$ & $2.17 \ignore{\pm 0.01}$&$0.854 \ignore{\pm 0.004}$ &$ 0.00\ignore{\pm 7}$\ignore{e-5} & $0.808 \ignore{\pm 0.001}$&$5.59 \ignore{\pm 0.01}$ & & & \cite{Adam:2016dau} \\
        $\text{K}^*(892)^0$                                      & $0.0 < \pt < \hspace{4pt}6.0$& $0.0 < \pt < \hspace{4pt}6.0$ &$0.303 \ignore{\pm 0.007}$ &$0.79 \ignore{\pm 0.02}$ &$0.00 \ignore{\pm 2}$\ignore{e-5} &$1.07 \ignore{\pm 0.01}$ &$5.37 \ignore{\pm 0.05}$ & & & \cite{Abelev:2012hy} \\
        $\text{p}/\overline{\text{p}}$                          & $0.3 < \pt < 20.0$ & $0.4 < \pt < 20.0$ &$0.604 \ignore{\pm 0.008}$ &$0.602 \ignore{\pm 0.007}$ &$0.00 \ignore{\pm 2}$\ignore{e-5} &$1.210 \ignore{\pm 0.005}$ &$6.86 \ignore{\pm 0.03}$ & & & \cite{Adam:2016dau} \\
        $\Omega^-/\overline{\Omega}^+$                          & $0.8 < \pt < \hspace{4pt}5.0$& $0.8 < \pt < \hspace{4pt}5.0$ &$0.0014 \ignore{\pm 0.0002}$ &$0.15 \ignore{\pm 0.02}$ &$0.00 \ignore{\pm 1}$\ignore{e-5} &$4.34 \ignore{\pm 0.45}$ &$ 15.2\ignore{\pm 1.7}$ & & & \cite{Abelev:2012jp} \\
        $\Xi^-/\overline{\Xi}^+$                                & $0.6 < \pt < \hspace{4pt}8.5$& $0.6 < \pt < \hspace{4pt}8.5$ &$0.012 \ignore{\pm 0.001}$ &$0.56 \ignore{\pm 0.03}$ &$0.00 \ignore{\pm 3}$\ignore{e-5} &$1.75 \ignore{\pm 0.02}$ &$7.1 \ignore{\pm 0.1}$ & & & \cite{Abelev:2014qqa} \\
        $\Sigma(1385)^{\pm}/\overline{\Sigma}(1385)^{\pm}$      & $0.7 < \pt < \hspace{4pt}6.0$& $0.7 < \pt < \hspace{4pt}6.0$ &$0.01 \ignore{\pm 2}$\ignore{e-5} &$0.76 \ignore{\pm 1}$\ignore{e-6} &$ 0.00\ignore{\pm 2.6}$ &$ 1.6\ignore{\pm 1}$\ignore{e-6} &$6.5 \ignore{\pm 3}$\ignore{e-11} & & & \cite{Abelev:2014qqa} \\
        $\omega(782)$      & $2.0 < \pt < 17.0$      & $2.0 < \pt < 17.0$ &$0.036 \ignore{\pm 2}$\ignore{e-5} &$0.707 \ignore{\pm 1}$\ignore{e-6} &$0.00 \ignore{\pm 2.6}$ &$ 2.01\ignore{\pm 1}$\ignore{e-6} &$6.00 \ignore{\pm 3}$\ignore{e-11} & & & \cite{Peresunko:2012tt}* \\
        \hline
        \mbox{}\\
        \hline
        Tsallis (\Eq{eq:tsallis}) && & $\text{d}N/\text{d}y$ & $n$ & $T$ & & & & & \\ 
        \hline 
        $\phi$                                                  & $0.4 < \pt < \hspace{4pt}6.0$& $0.4 < \pt < \hspace{4pt}6.0$ &$ 0.197\ignore{\pm 0.002}$ &$6.85 \ignore{\pm 0.17}$ &$0.277 \ignore{\pm 0.004}$ & & & & & \cite{Abelev:2012hy} \\
        \hline
        \mbox{}\\
        \hline
        Empirical ratio (\Eq{eq:softHard}) && & $A$ & $\beta$ & $T$ & $N$ & $B$ & $p_0$ & $n$ & \\ 
        \hline 
        $\eta/\pi^{0}$                                          & $0.4 < \pt < 15.0 $ & $0.4 < \pt < \hspace{4pt}8.0 $ &$0.55$ & $0.99985$ &$11.4$ &$-233.4$ &$0.00036$ &$-37.54$ &$130.3$ & \cite{Abelev:2012cn} \\
        $\rho^{0}/\pi^{\pm}$                                & $0.5 < \pt < 12.0 $& $0.5 < \pt < 12.0 $ & 0.85&0.667 &10.67 &94.75 &-0.0099 &-1.19 & 1.36& \cite{Riabov:2017jig}** \\      
                \hline
    \end{tabular}
    \caption{Summary of fit parameters obtained for the particle spectra shown in \Figs{fig:yieldsAndFits}{fig:ratiosYieldsToFits}. 
             Fit parameter uncertainties are not given since the parameterizations are just used to approximate the data without implying any physical interpretation. 
             (*) For the $\omega(782)$ meson, the preliminary measurement \cite{Peresunko:2012tt} has been used for the parameterization. 
             (**) The $\rho^{0}$ and $\rho^{\pm}$ spectra were obtained by multiplying the parameterization of the preliminary $\rho^{0}/\pi^{\pm}$ from pp $\sqrt{s}=2.76$ TeV \cite{Riabov:2017jig} with the given charged pion parameterization. }
    \label{tab:particlesAndParameters}
\end{table}
\end{minipage}\vspace{0.7cm}
\begin{minipage}{0.491\textwidth}
  \hspace{-0.7cm}
  \includegraphics[width=1.05\textwidth]{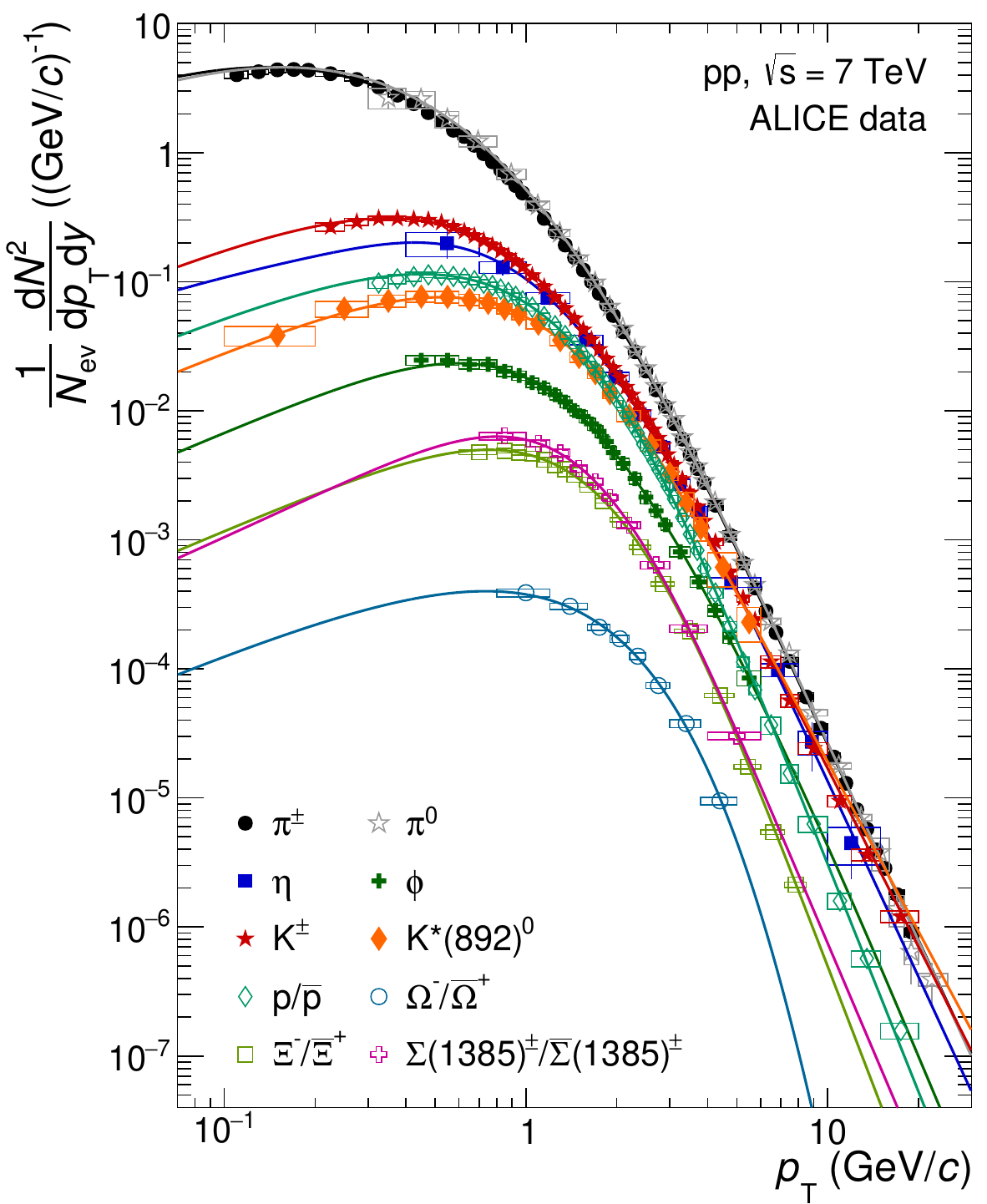}
  \caption{Identified, $\pt$ differential particle yields measured by ALICE in \pp\ collisions at $\sqrt{s}=7$~TeV and corresponding parametrizations. 
           Statistical uncertainties are shown as error bars, systematic uncertainties are shown as open boxes. 
           The shown data are $\pi^{\pm}$ \cite{Adam:2016dau}, $\pi^0$ \cite{Abelev:2012cn}, $\eta$ \cite{Abelev:2012cn}, $\phi$ \cite{Abelev:2012hy}, $\text{K}^{\pm}$ \cite{Adam:2016dau}, $\text{K}^{*}(892)^{0}$ \cite{Abelev:2012hy}, $\text{p}/\bar{\text{p}}$ \cite{Adam:2016dau}, $\Omega^-/\overline{\Omega}^+$ \cite{Abelev:2012jp}, $\Xi^-/\overline{\Xi}^+$ \cite{Abelev:2014qqa} and $\Sigma(1385)^{\pm}/\overline{\Sigma}(1385)^{\pm}$ \cite{Abelev:2014qqa}. 
           The lines are fits to the data summarized in \Tab{tab:particlesAndParameters}.}
  \label{fig:yieldsAndFits}
\end{minipage} \hspace{0.1cm}
\begin{minipage}{0.491\textwidth}
  \vspace{-1.1cm}
  \hspace{-0.52cm}
  \includegraphics[width=1.05\textwidth]{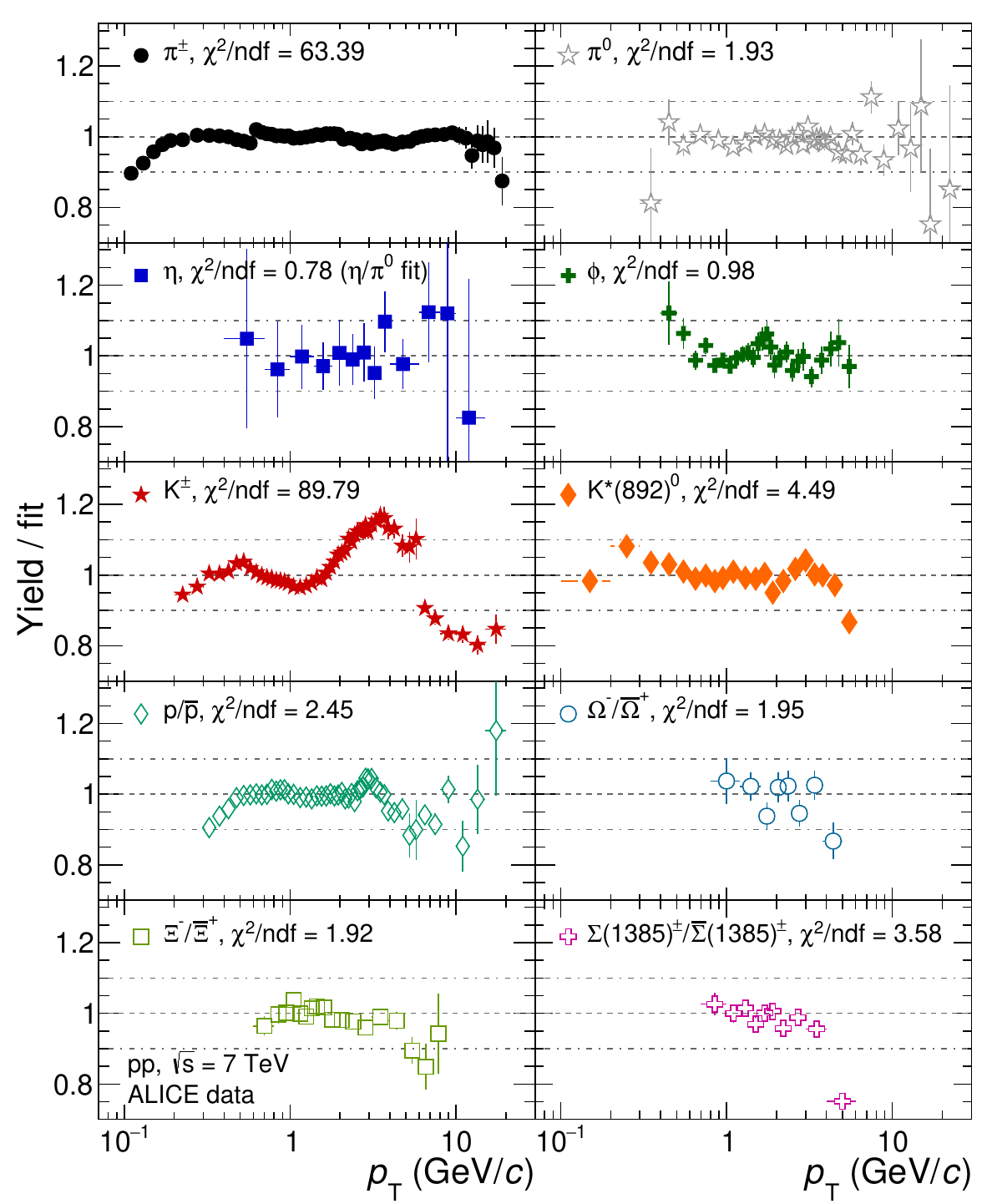}
  \caption{Ratios of measured yields to the respective parametrizations and reduced $\chi^2$ values of the fits. In the case of the $\eta$ meson, the reduced $\chi^2$ value of the corresponding $\eta/\pi^0$ ratio fit is given. Only statistical uncertainties are shown. For more info see caption of \Fig{fig:yieldsAndFits}.}
  \label{fig:ratiosYieldsToFits}
\end{minipage}
\end{figure*}

A large part of the available identified and $\pt$ differential particle yields measured by the ALICE experiment in proton-proton collisions at $\sqrt{s}=7$~TeV are summarized in \Fig{fig:yieldsAndFits}, where statistical uncertainties are displayed as error bars and systematic uncertainties as open boxes. 
The particles shown are $\pi^{\pm}$ \cite{Adam:2016dau}, $\pi^0$ \cite{Abelev:2012cn}, $\eta$ \cite{Abelev:2012cn}, $\phi$ \cite{Abelev:2012hy}, $\text{K}^{\pm}$ \cite{Adam:2016dau}, $\text{K}^{*}(892)^{0}$ \cite{Abelev:2012hy}, $\text{p}/\bar{\text{p}}$ \cite{Adam:2016dau}, $\Omega^-/\overline{\Omega}^+$ \cite{Abelev:2012jp}, $\Xi^-/\overline{\Xi}^+$ \cite{Abelev:2014qqa} and $\Sigma(1385)^{\pm}/\overline{\Sigma}(1385)^{\pm}$ \cite{Abelev:2014qqa}. 
In the case of charged particles, the average of all contributing yields (e.g.\ $\pi^{\pm} = (\pi^+ + \pi^-)/2$) is used synonymously for the corresponding species throughout this article. 
Furthermore, parametrizations fitted to the different yields are also shown in \Fig{fig:yieldsAndFits}.
The ratios between the data points and the fits are presented in \Fig{fig:ratiosYieldsToFits}. 
Systematic uncertainties were not included in the fit and hence are not shown.
Except for $\text{K}^{\pm}$, the fits typically describe the data to better than $10$\%.
The fit parameters are summarized in \Tab{tab:particlesAndParameters} and are explained below.

All particle yields, except those of the $\eta$ and $\phi$ mesons, were parametrized with a modified Hagedorn function~\cite{Khandai:2012xx,Adare:2009qk}, multiplied with $\pt$ to account for the spectra being non-invariant:
\begin{align}
	\frac{\text{d}^2N}{\text{d}y\text{d}\pt} = \pt \times A \times \left( \exp{\left( a\pt + b\pt^2 \right)} + \frac{\pt}{p_0} \right)^{-n}.
	\label{eq:modifiedHagedorn}
\end{align}
This functional form approaches an exponential at low $\pt$ and a power law at larger transverse momenta and describes the measured spectra very well over the full measured range in $\pt$. 
In the case of the $\text{K}^{\pm}$ meson, a step at $\pt\approx6$~GeV/$c$ is observed, which can be traced back to significantly different analysis techniques used for the lower and higher parts of the measured spectrum~\cite{Adam:2016dau}.

The yield of the $\phi$ meson was best described by a Tsallis function \cite{Cleymans:2012ya} multiplied with $\pt$, which can be written as
\begin{align}
	\frac{\text{d}^2N}{\text{d}y\text{d}\pt} = \pt \times \frac{\text{d}N}{\text{d}y} \times \frac{(n - 1)(n - 2)}{nT (nT + m ( n - 2 ) )}\left( 1 + \frac{\mt - m}{nT} \right)^{-n}.
	\label{eq:tsallis}
\end{align}

\begin{figure}[t!]
  \includegraphics[width=0.49\textwidth]{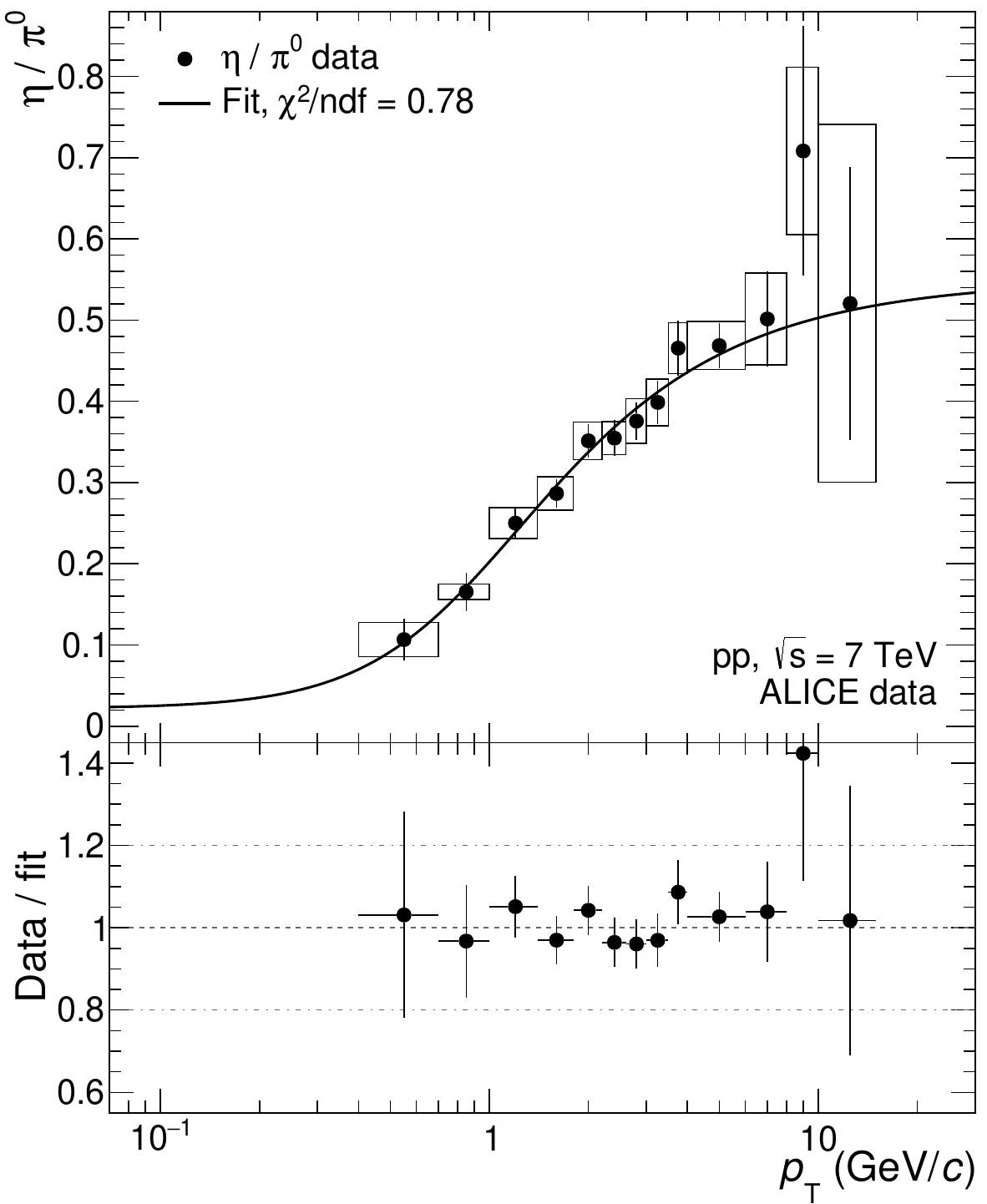}\\ \vspace{-0.3cm}
  \caption{The $\eta/\pi^0$ ratio measured in \pp\ collisions at $\sqrt{s}=7$~TeV~\cite{Abelev:2012cn}, the corresponding parametrization and the ratio between the data and fit. 
           Statistical uncertainties are shown as error bars, systematic uncertainties are shown as open boxes.}
  \label{fig:etaToPi0Ratio}
\vspace{0.7cm}
  \includegraphics[width=0.49\textwidth]{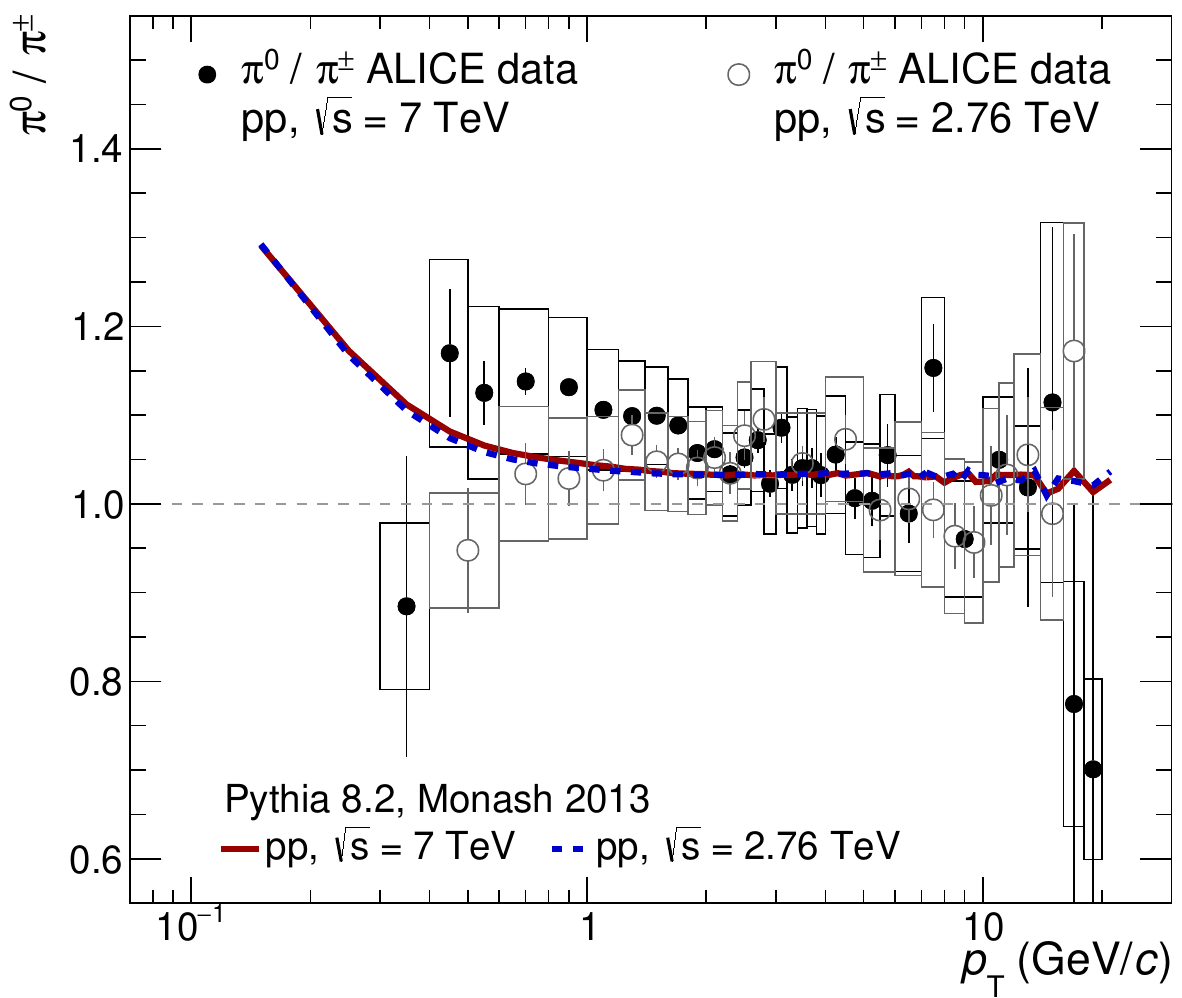}
  \caption{The $\pi^0/\pi^\pm$ ratio measured in \pp\ collisions at $\sqrt{s}=7$~TeV \cite{Abelev:2012cn,Abelev:2014laa} and $\sqrt{s}=2.76$~TeV \cite{Acharya:2017hyu,Abelev:2014laa}. 
           Statistical uncertainties are shown as error bars, systematic uncertainties are shown as open boxes. 
           The ratios are compared to calculations using PYTHIA 8.2, Monash 2013 tune~\cite{Sjostrand:2014zea}.}
  \label{fig:neutralToChargedPi}
\end{figure}

A different approach for the fitting procedure was chosen for the $\eta$ yield. 
In order to reliably describe the high $\pt$ part of the spectrum, the ratio to the neutral pion yield, shown in \Fig{fig:etaToPi0Ratio}, was fitted first with an empirical~\cite{Klaus} function 
\begin{align} 
	\frac{\eta}{\pi^0}(\pt) = \frac{ A \cdot \exp \left( \frac{ \beta \pt - \mt^{\eta} }{ T \sqrt{1 - \beta^2} } \right) + N  \cdot B \cdot \left( 1 + \left( \frac{\pt}{p_0} \right)^2 \right)^{-n} }{ \exp \left( \frac{ \beta \pt - \mt^{\pi^0} }{ T \sqrt{1 - \beta^2} } \right) + B \cdot \left( 1 + \left( \frac{\pt}{p_0} \right)^2 \right)^{-n} }.
	\label{eq:softHard}
\end{align}
This function contains two separate contributions from soft and hard processes, where $B$ is a relative normalization between the soft and hard part of the parametrization and $N$ is the constant ratio between the two particle species which is approached at high $\pt$. 
The soft part of the spectra is described by a blast wave inspired function~\cite{Schnedermann:1993ws} that depends on the radial flow velocity $\beta$ and the kinetic freeze-out temperature $T$. 
While this Ansatz was originally intended to describe the production in a system where a hot medium is produced, it also yields a good description of the data in proton-proton collisions. 
The fit to the $\eta/\pi^0$ ratio is compared with the data in \Fig{fig:etaToPi0Ratio}. 
At high $\pt$, the fit approaches a constant ratio of $R_{\eta/\pi^0} \approx 0.55$, which is larger than the global average of $R_{\eta/\pi^0} = 0.459 \pm 0.013$ at LHC energies~\cite{Acharya:2017tlv}. 
However, the fit is only intended to serve as the best possible parametrization of the available data points within the measured range and should not be taken as a quantitative estimate of the $\eta/\pi^0$ ratio at high $\pt$.
The parametrization of the $\eta$ yield is then obtained by multiplying the parametrization of the ratio with the parametrization of the neutral pion yield. 
It should be noted that the reduced $\chi^2$ value for the $\eta$ meson quoted in \Fig{fig:ratiosYieldsToFits} refers to the fit of the particle ratio shown in \Eq{eq:softHard}.

Although typically neutral pions are used as a basis for transverse mass scaling of mesons, we use the charged pions since they are measured with a better precision and to lower $\pt$ as can be seen in \Fig{fig:yieldsAndFits}. 
The direct comparison of the corresponding neutral and charged pion spectra for $\sqrt{s}=7$~TeV \cite{Abelev:2012cn,Abelev:2014laa} and $\sqrt{s}=2.76$~TeV \cite{Acharya:2017hyu,Abelev:2014laa} is shown in \Fig{fig:neutralToChargedPi}. 
The statistical and systematic uncertainties are shown as error bars and open boxes, respectively.
As can be seen, the charged and neutral pions spectra agree within the systematic and statistical uncertainties, in particular when considering that part of the systematic uncertainties is correlated in $\pt$. 
The difference at low $\pt$ is expected from isospin violating decays, and the ratios are consistent with predictions from PYTHIA~\cite{Sjostrand:2014zea}.

\begin{figure}[t!]
	\includegraphics[width=0.49\textwidth]{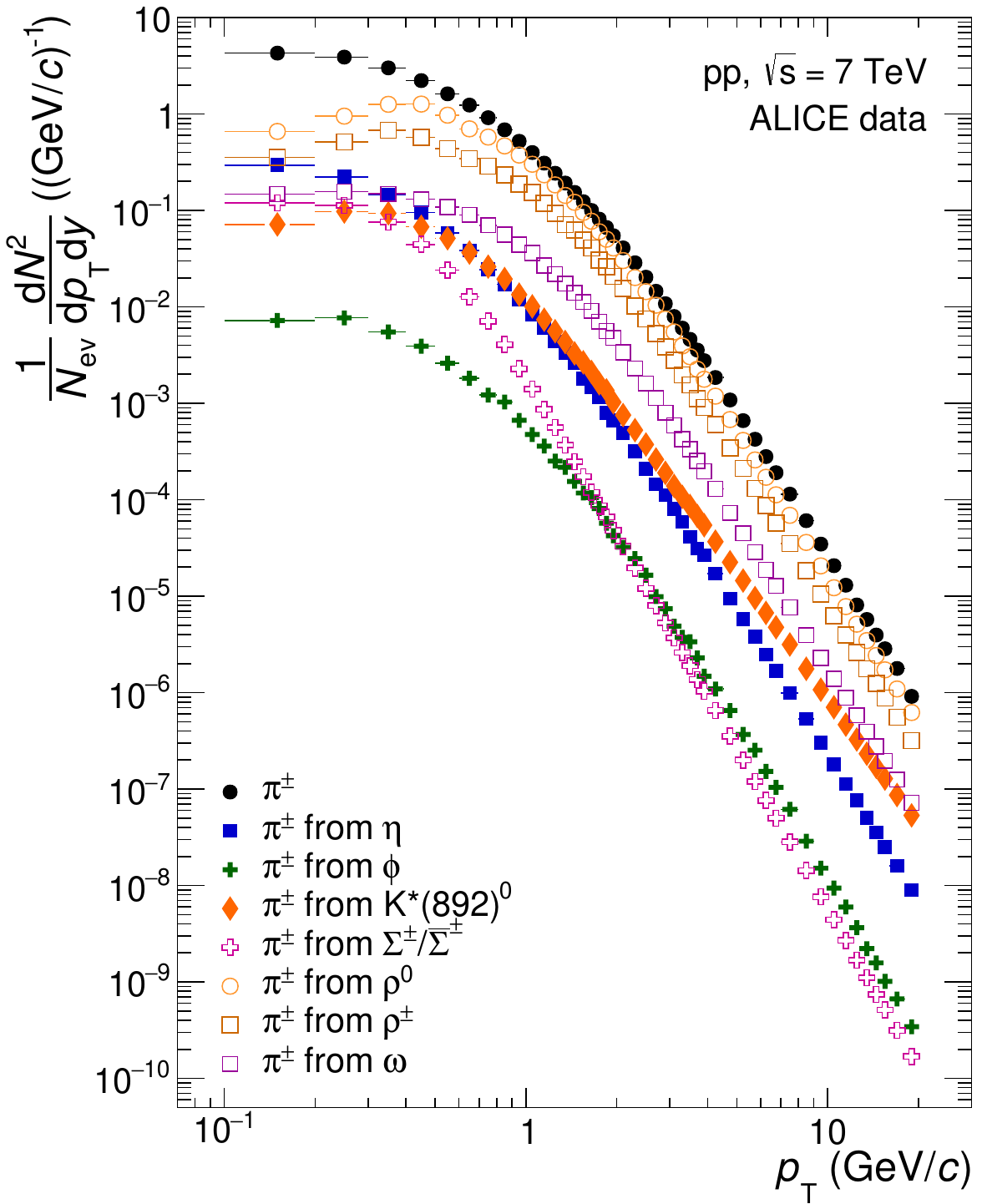}\\\vspace{-0.3cm}
	\caption{Primary charged pion yields (black) in \pp\ collisions at $\sqrt{s}=7$~TeV , according to the primary particle definition in \cite{ALICE-PUBLIC-2017-005}, generated according to the charged pion parameterization given in \Tab{tab:particlesAndParameters}. Furthermore, the different feed-down contribution from a particle decay simulation are shown per source. The uncertainties given are related to the statistics of the decay particle simulation and the uncertainties on the parameterization of the sources are not taken into account.}
	\label{fig:chargedPionYields}
\end{figure}

The general transverse mass scaling relation as represented in \Eq{eq:scalingRelation} is defined for primary particles. 
However, there is a significant contribution to the measured yields from particle decays, especially to light particles as $\pi^{\pm}$~(or $\pi^0$). 
For example, in the case of the $\pi^{\pm}$ measurement, contributions from weak decays (so-called secondaries) are subtracted from the measured spectrum \cite{Adam:2016dau} in accordance with the definition of primary particles that is used within ALICE~\cite{ALICE-PUBLIC-2017-005}. \\
\indent Nevertheless, this does not remove all possible feed-down contributions from resonance decays. 
In order to estimate the effect of feed-down on the applicability of the scaling relation, the charged pion yield was calculated for some of those contributions. 
This was done with a particle decay simulation using the PYTHIA 6.4 ``particle-decayer'' algorithm~\cite{Sjostrand:2006za}.
The largest sources of charged pions are $\rho^{0,\pm}$, $\omega$, $\eta$, $\eta'$, $\phi$, $\text{K}^{*}(892)^{0}$ as well as the various $\Delta$, $\Xi$, $\Omega$ and $\Sigma$ resonances. 
In \pp\ collisions at $\sqrt{s}=7$~TeV the $\eta$~\cite{Abelev:2012cn}, $\phi$~\cite{Abelev:2012hy}, $\text{K}^{*}(892)^{0}$~\cite{Abelev:2012hy} and $\Sigma(1385)^{\pm}$~\cite{Abelev:2014qqa} $\pt$-differential cross sections have been measured and a preliminary $\omega$ yield \cite{Peresunko:2012tt} above $4$~GeV/$c$ is available.
For the $\rho$ meson no yield measurements in \pp\ collisions at $\sqrt{s}=7$~TeV have been performed, yet.
Thus, the preliminary $\rho^0/\pi^\pm$ for \pp\ collisions at $\sqrt{s}=2.76$~TeV \cite{Riabov:2017jig} has been parametrized, assuming no collision energy dependence of the particle ratio. 
Afterwards, this has been multiplied with the charged pion yield seen in \pp\ collisions at $\sqrt{s}=7$~TeV to obtain the $\pt$ differential $\rho$ yield at this center-of-mass energy.
The remaining particles were generated according to the parametrized $\pt$-yields quoted in \Tab{tab:particlesAndParameters}
uniformly in rapidity and azimuthal angle with the ranges $[-1, 1]$ and $[0, 2\pi]$, respectively. 
Charged pions from decays of the generated mother particles were counted within the originally analyzed rapidity range of $[-0.8, 0.8]$. 
The contributions obtained from the decay simulation as well as the primary charged pion spectrum, which has been generated according to the measured spectrum, are shown in \Fig{fig:chargedPionYields}. 
It can be seen that the by far strongest contribution among the considered sources arises from $\rho$ decays followed by the $\omega$ decays, while the others are sub-dominant.
Above transverse momenta of $3$--$4$~GeV/$c$ all feed-down contributions appear to have a similar slope, which is expected as all mother particles have a similar slope as well.
Below $3$~GeV/$c$, on the other hand, the shape of the different contributions differs significantly and the total charged pion yield can be understood as the sum of all of those contributions.
However, the uncertainties of the parameterizations of the mother particles are significant below $1$~GeV/$c$ and thus the subtraction of the different feed-down contributions has not been done.
Additional contributions will arise from the remaining baryonic resonances, which should have similar transverse momentum distribution as the contribution from the $\Sigma(1385)^{\pm}$ decays, but are not considered in this study.\\
\indent Based on these observations it can be inferred, that the transverse mass scaling based on the charged or neutral pion will break down at low $\pt$, as the measured pion spectrum is not governed by just a single (exponential) function of the pion mass.
For \pp\ collisions at center-of-mass energies $8$--$16$ GeV, for which the initial scaling relation was observed, heavy ---in particular baryonic--- resonance production is strongly suppressed and thus the pion spectra from different sources are more similar.
With increasing center-of-mass energy, however, the contribution increases changing the shape of the pion spectrum at low transverse momenta and thus the scaling only holds at higher transverse momenta, where the spectra are dominated by the hard particle production mechanisms. 
If a heavier reference particle is chosen for the scaling the influence of the resonances decreases.
However, the scaling might also break down at lower $\pt$ than for the pions for the same reason and a similar feed-down study should be performed for the corresponding reference particles.

\section{Results}
\label{sec:results}
\begin{figure*}
  \begin{minipage}{\textwidth}
    \center
    \includegraphics[width=0.49\textwidth]{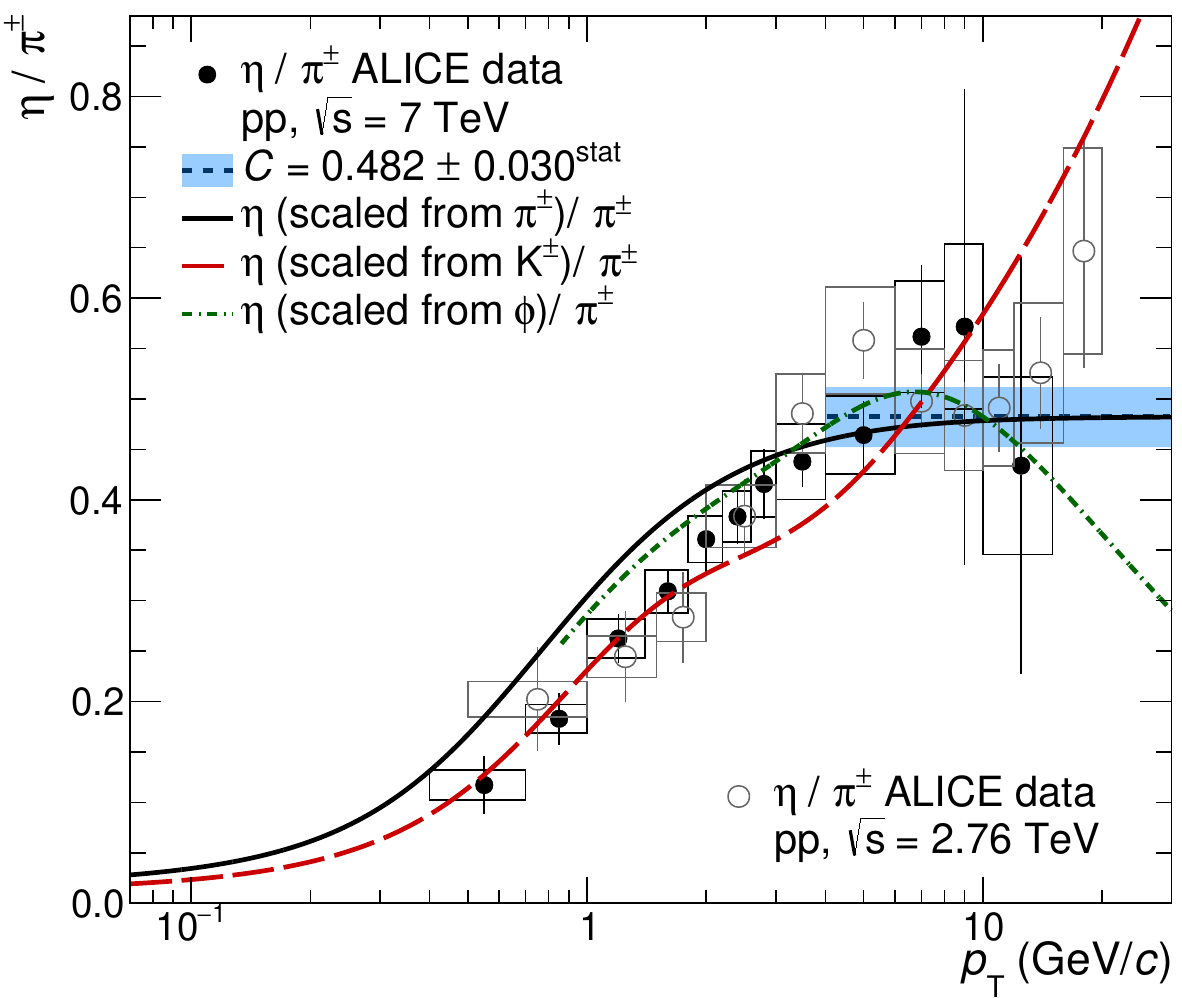}
    \includegraphics[width=0.49\textwidth]{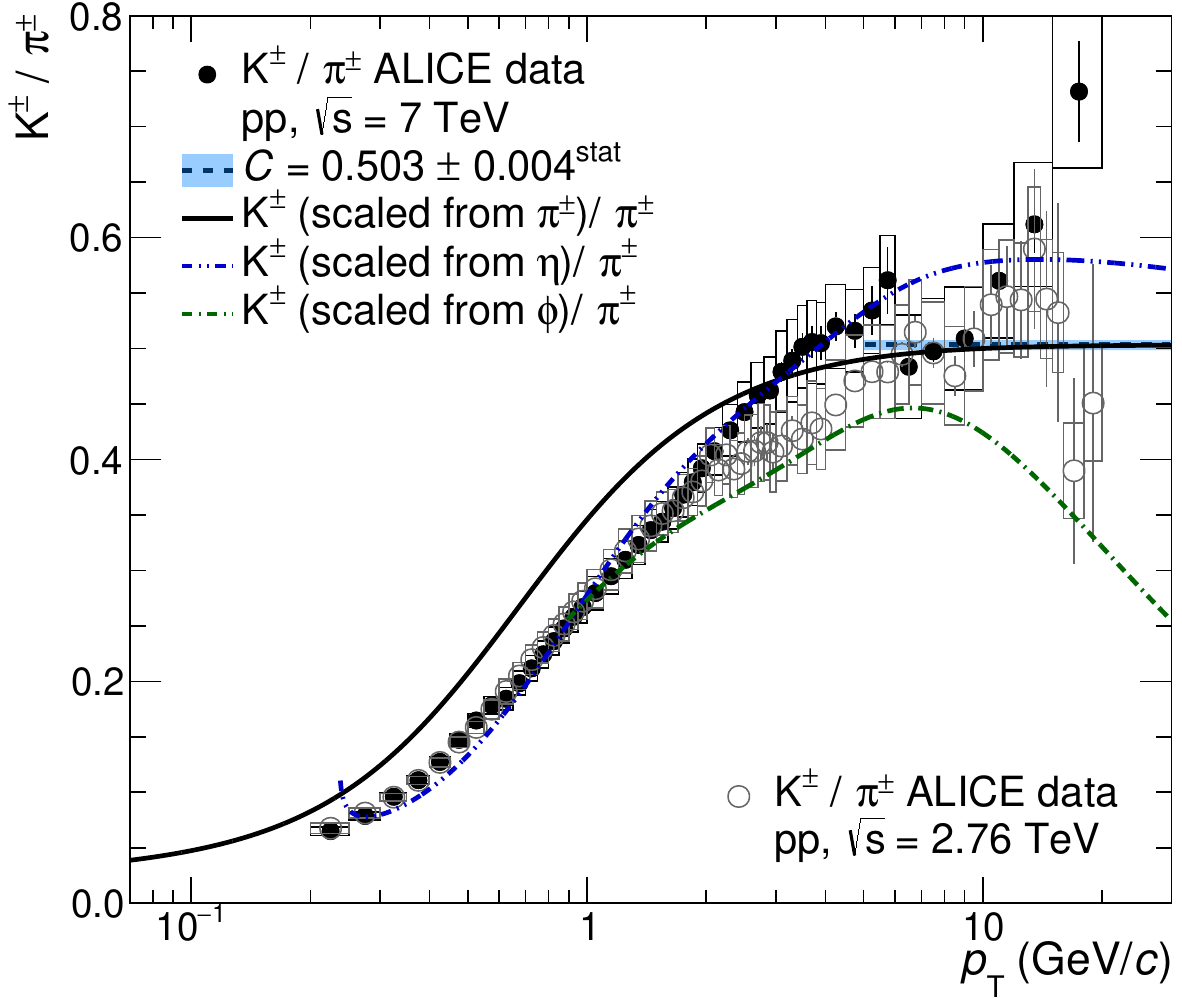}  \\\vspace{-0.3cm}
    \caption{The $\eta/\pi^\pm$ (left) and K$^\pm/\pi^\pm$ ratio (right) measured in \pp\ collisions at $\sqrt{s}=7$~TeV together with the ratios using the different particles as basis for the scaling of the $\eta$ meson and charged kaon and the parameterization of the charged pion, respectively. The fitted high $\pt$ constant $C$ is reported as well as a blue band in the $\pt$ region, where it was obtained. For comparison the corresponding particle ratios measured in \pp\ collisions at $\sqrt{s}=2.76$~TeV are displayed as well \cite{Acharya:2017hyu,Abelev:2014laa}.}
    \label{fig:ratios1}
    \vspace{0.3cm}
    \includegraphics[width=0.49\textwidth]{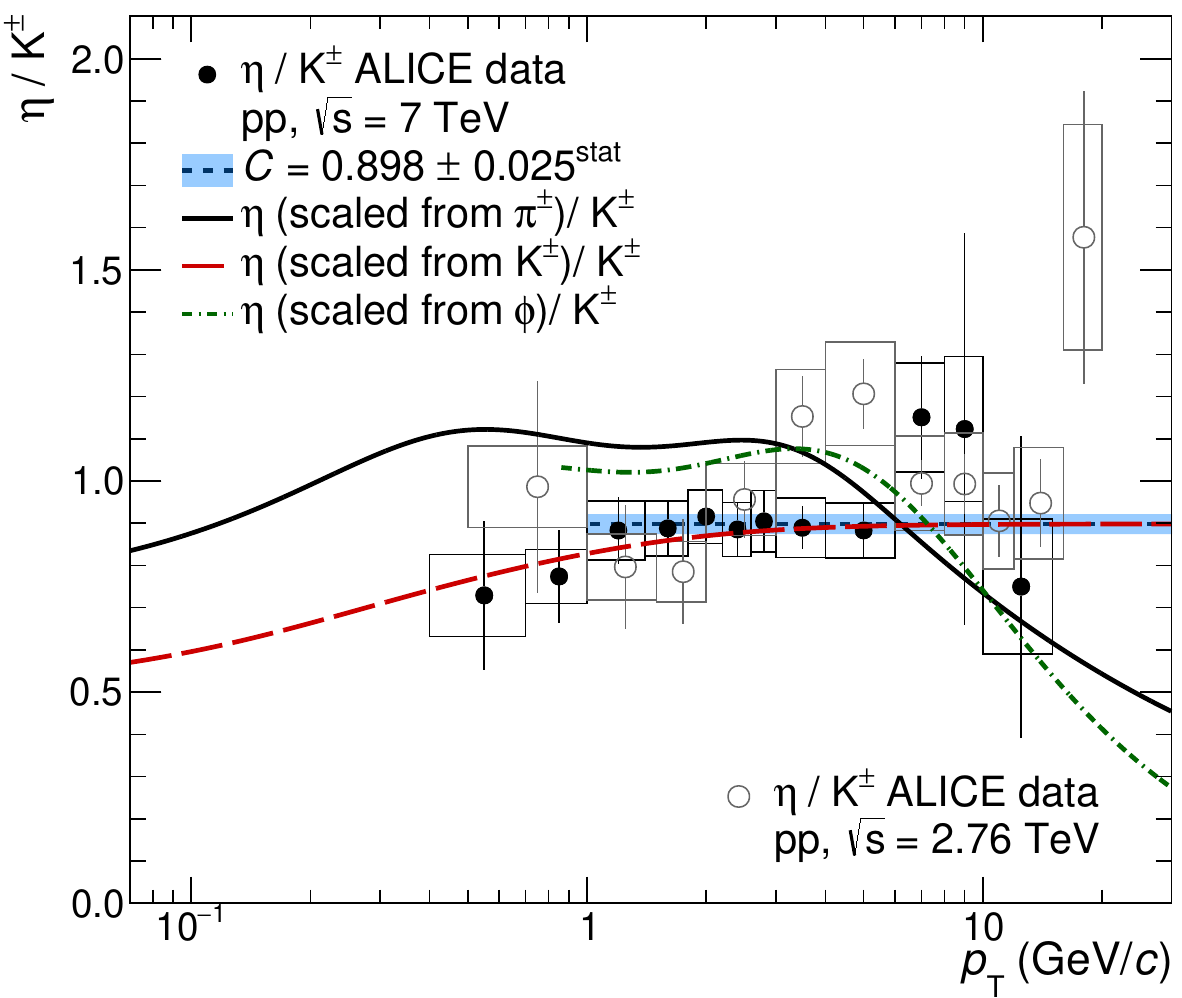}
    \includegraphics[width=0.49\textwidth]{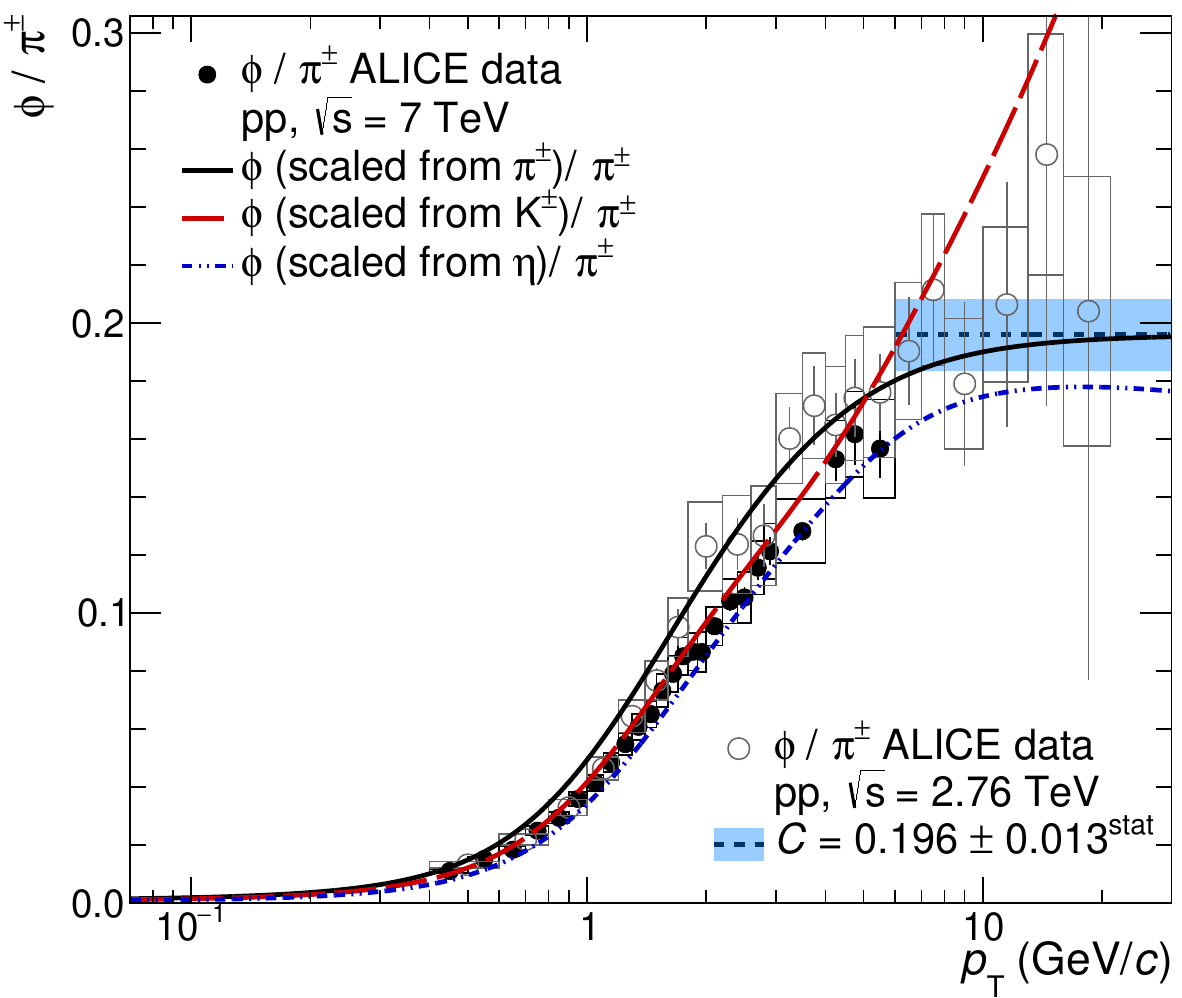}\\ \vspace{-0.3cm}
    \caption{The $\eta/$K$^\pm$ (left) and $\phi/\pi^\pm$ ratio (right) measured in \pp\ collisions at $\sqrt{s}=7$~TeV together with the ratios using the different particles as basis for the scaling of the $\eta$ and $\phi$ meson and the parameterizations of the charged kaon and pion, respectively. For comparison the corresponding particle ratios measured in \pp\ collisions at $\sqrt{s}=2.76$~TeV are displayed as well \cite{Acharya:2017hyu,Abelev:2014laa,Adam:2017zbf}. The fitted high $\pt$ constant $C$ is reported as well as a blue band in the $\pt$ region, where it was obtained. The scaling constant for the $\phi/\pi^\pm$ is obtained solely on the data obtained for \pp\ collisions at $\sqrt{s}=2.76$~TeV , as the plateau value has not been reached for the results at $7$~TeV. }
    \label{fig:ratios2}
  \end{minipage}
\end{figure*}

\begin{figure*}
  \begin{minipage}{\textwidth}
    \center
    \includegraphics[width=0.49\textwidth]{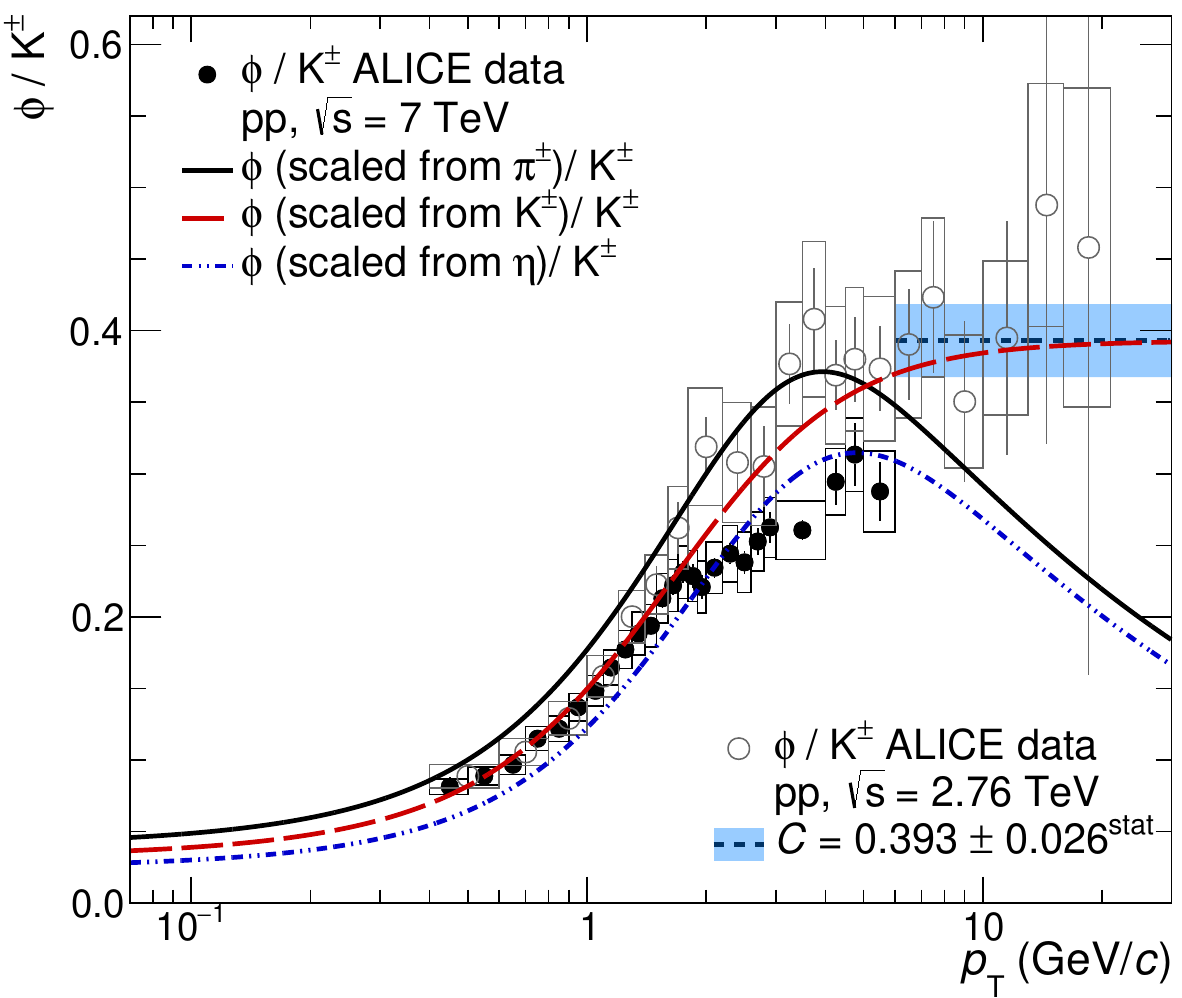}
    \includegraphics[width=0.49\textwidth]{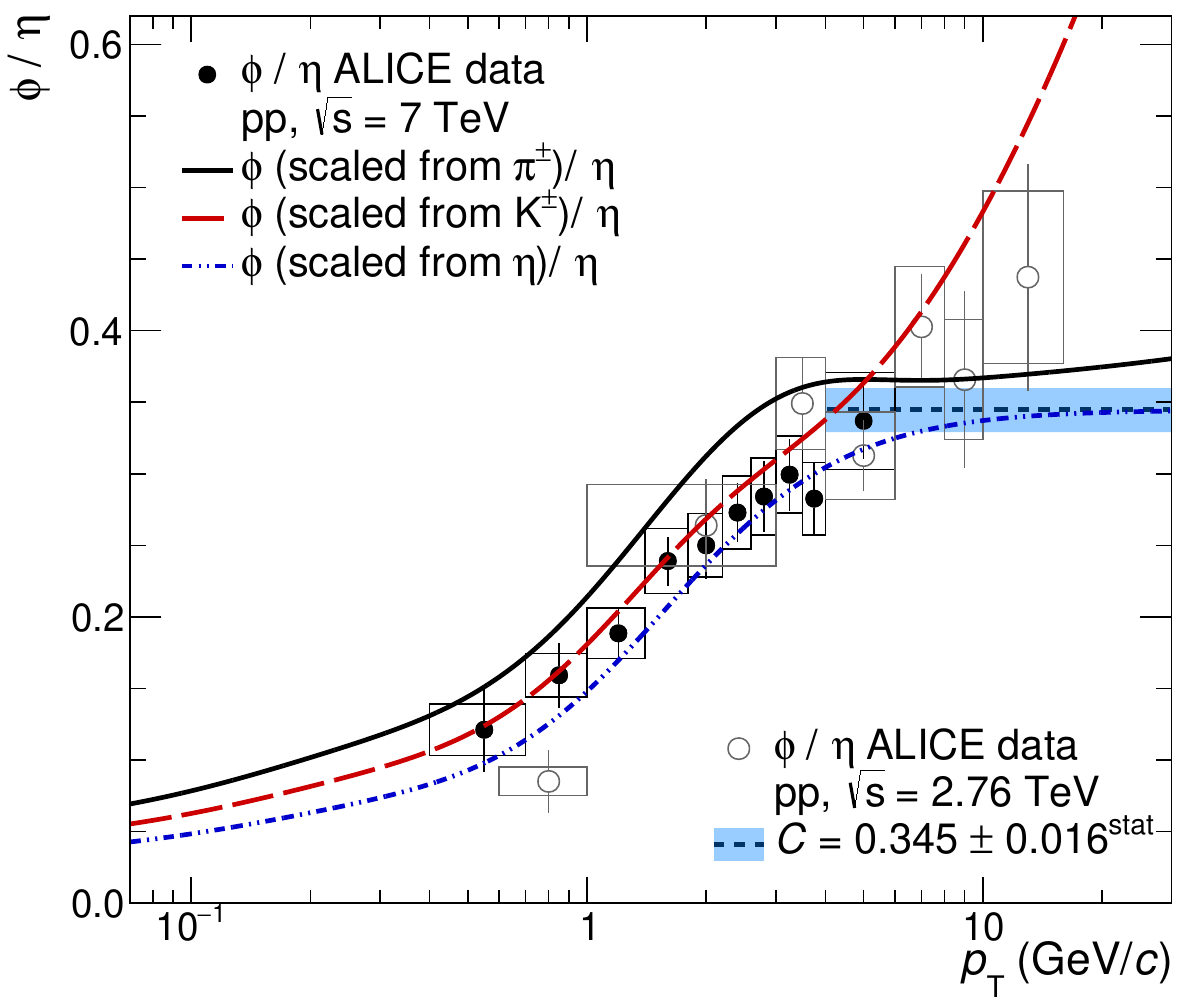}\\\vspace{-0.3cm}
    \caption{The $\phi/$K$^\pm$ (left) and $\phi/\eta$ ratio (right) measured in \pp\ collisions at $\sqrt{s}=7$~TeV together with the ratios using the different particles as basis for the scaling of the $\phi$ meson and the parameterizations of the charged kaon and $\eta$ meson, respectively. For comparison the corresponding particle ratios measured in \pp\ collisions at $\sqrt{s}=2.76$~TeV are displayed as well \cite{Acharya:2017hyu,Abelev:2014laa,Adam:2017zbf}. The fitted high $\pt$ constant $C$ is reported as well as a blue band in the $\pt$ region, where it was obtained. Both scaling constants mainly rely on the data obtained for \pp\ collisions at $\sqrt{s}=2.76$~TeV, as the plateau value has not been reached for the results at $7$~TeV.}
    \label{fig:ratios3}
    \vspace{0.3cm}
    \includegraphics[width=0.49\textwidth]{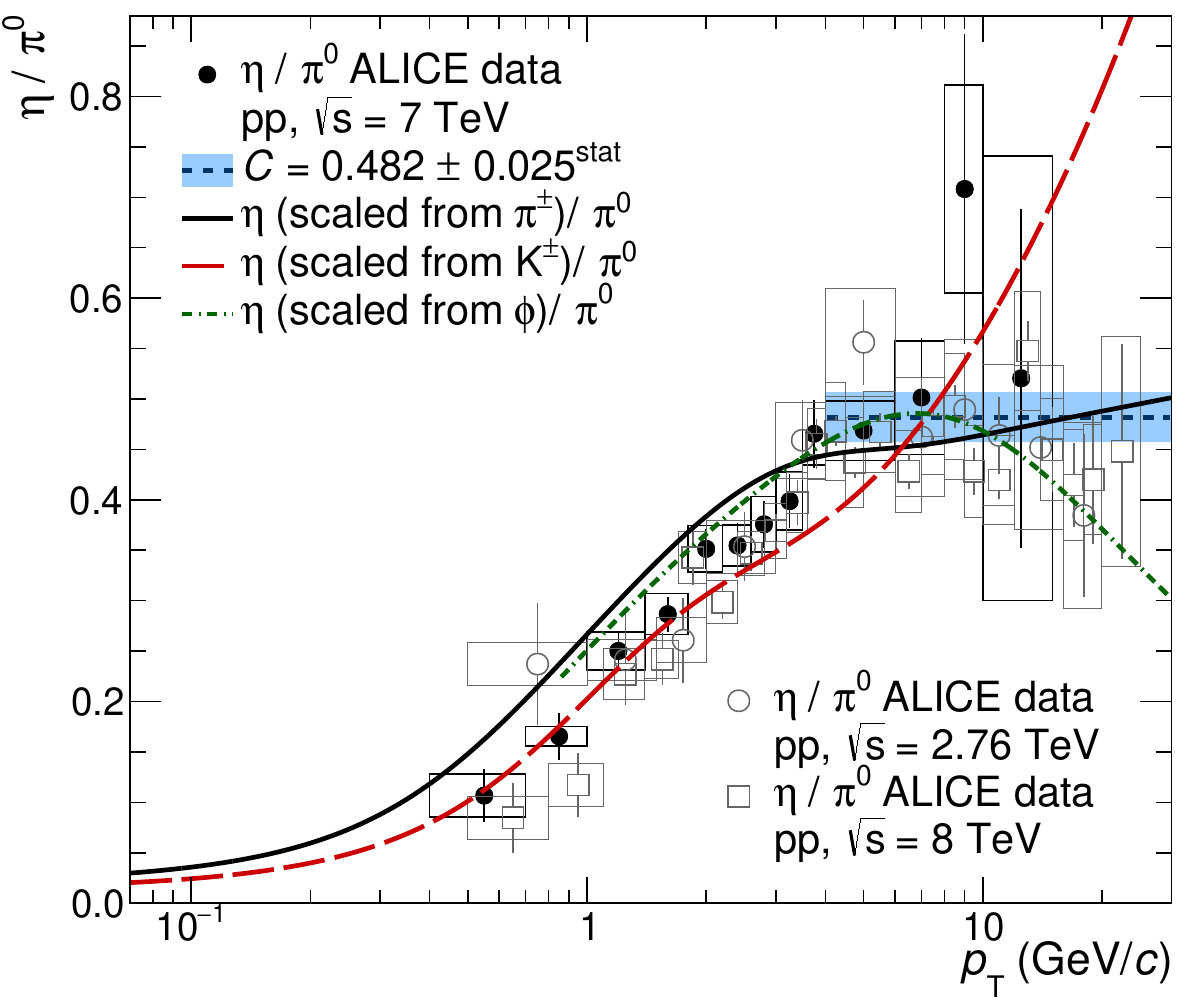}
    \includegraphics[width=0.49\textwidth]{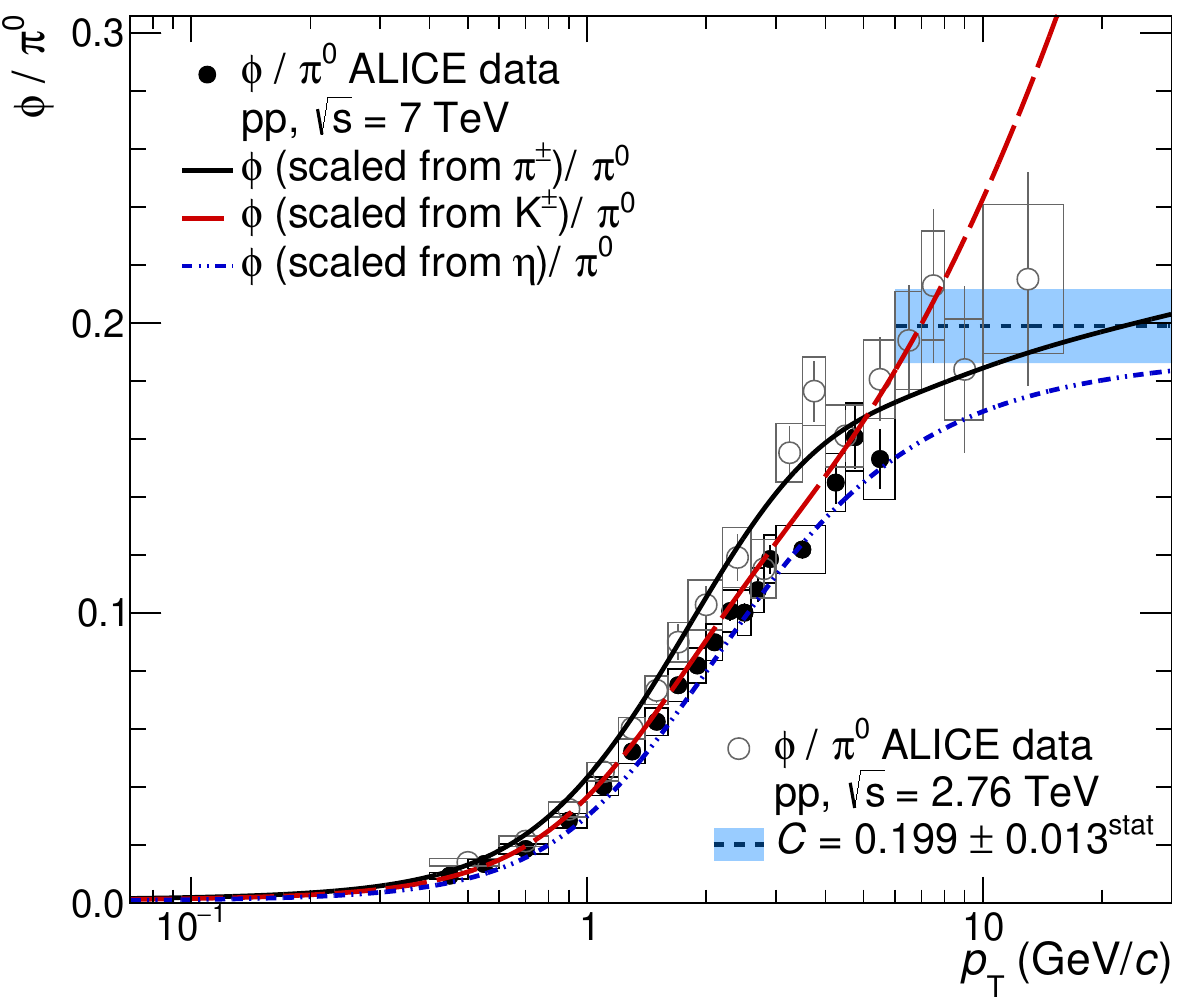}\\\vspace{-0.3cm}
    \caption{The $\eta/\pi^0$ (left) and $\phi/\pi^0$ ratio (right) measured in \pp\ collisions at $\sqrt{s}=7$~TeV together with the ratios using the different particles as basis for the scaling of the $\eta$ and $\phi$ meson, respectively, and the parameterization of the neutral pions. For comparison the corresponding particle ratios measured in \pp\ collisions at $\sqrt{s}=2.76$~TeV and $8$~TeV are displayed as well \cite{Acharya:2017hyu,Adam:2017zbf,Acharya:2017tlv}. The fitted high $\pt$ constant $C$ is reported as well as a blue band in the $\pt$ region, where it was obtained. For the $\phi/\pi^0$ the scaling constant was obtained based on the data measured at $\sqrt{s}=2.76$~TeV, as the plateau value has not been reached for the results at $7$~TeV.}
    \label{fig:ratioToNPi}

  \end{minipage}
\end{figure*}

The general transverse mass scaling relation according to \Eq{eq:scalingRelation} was applied to the yield parametrizations of different reference particles:\ $\pi^{\pm}$, $\eta$, $\phi$ and $\text{K}^{\pm}$, in order to evaluate the applicability of the scaling.
The corresponding scaling parameter $C$ from \Eq{eq:scalingRelation} has been estimated by fitting the respective particle ratios at high transverse momenta, where the $\pt$ threshold has been chosen such, that the asymptotic value has been reached.
For the ratios of the $\phi$ meson to the different reference particles the plateau values in \pp\ collisions at $\sqrt{s}=7$~TeV are not yet reached, thus the corresponding ratios in \pp\ collisions at $\sqrt{s}=2.76$~TeV have been fitted, as they extend further in $\pt$.
This assumes that the $\phi$-related ratios at high $\pt$ do not depend strongly on the center-of-mass energy at LHC energies.
The corresponding particle ratios with the various reference particles can be found in \Figsm{fig:ratios1}{fig:ratios3}, where the obtained high $\pt$ values are displayed as blue band including the corresponding uncertainties.
Furthermore, the particle ratios of the $\eta$ and $\phi$ meson with respect to the neutral pion are displayed in \Fig{fig:ratioToNPi} for completeness, as these are usually used for the validation of the $\mt$-scaling procedure for the direct photon and low mass di-electron analyses.
The obtained scaling parameters are listed in \Tab{tab:scalingParameters}. 

\begin{table}[t!]
  \setlength\tabcolsep{5pt}
  \begin{tabular}{l | c c c}
  \hline
                                    & \multicolumn{3}{c}{Scaled}                                                           \\
Basis                               &  $\eta$               & $\phi$                        & $(\text{K}^++\text{K}^-)/2$   \\
\hline
$\pi^0$                             & $0.482 \pm 0.025$     & $0.199 \pm 0.013$             & -                             \\
$(\pi^++\pi^{-})/2$                  & $0.482 \pm 0.030$     & $0.196 \pm 0.013$             & $0.503 \pm 0.004$             \\
$\eta$                              & -                     & $0.344 \pm 0.016$             & $1.113 \pm 0.031$             \\
$\phi$                              & $2.90 \pm 0.13$       &  -                            & $2.54 \pm 0.17$               \\
$(\text{K}^++\text{K}^-)/2$         & $0.898 \pm 0.025$     & $0.393 \pm 0.026$             & - \\ \hline
  \end{tabular} 
  \caption{Constant scaling factors ($C$ in the scaling relation, \Eq{eq:scalingRelation}) obtained from the ratio fits for different combinations of reference and scaled particle types for \pp\ collisions at $\sqrt{s}=7$~TeV at high transverse momenta. For the $\phi$ meson ratios the high $\pt$ points for \pp\ collisions at $\sqrt{s}=2.76$~TeV have been considered in the fits. }
  \label{tab:scalingParameters}
\end{table}


As the particle rest masses of the $\eta$ meson and charged kaon are fairly similar the scaling factor is very close to unity for these combinations, and the $\pt$ threshold for the high momentum fit can be significantly lowered from about $4$~GeV/$c$, as for most of the other combinations, to $1$~GeV/$c$.
In the case of the scaling of the $\eta$ and K$^{\pm}$ from the $\phi$ meson, large uncertainties on the scaling factors are observed. 
This is due to the large statistical uncertainties associated to the $\eta$ and $\phi$ spectra (see \Fig{fig:ratios3}) in combination with the strong restriction on the fit range due to the large mass of the $\phi$ meson. 
It should also be noted that some of the factors listed in \Tab{tab:scalingParameters} are larger than one, which is due to the fact that in those cases the reference particle is heavier than the particle that it is scaled to. 
This poses a limit for the scaled function since $\sqrt{m_{0,{\rm X}}^2 + p_{\text{T,X}}^2 - m_{0,{\rm R}}^2}$ becomes imaginary below a certain $\pt$.
In these cases a lower limit of $\pt \geq \sqrt{m_{0,{\rm R}}^2 - m_{0,{\rm X}}^2 }$ was used generally, i.e.\ also for the extrapolated scaled function. 
However, this limit does not affect the scaling procedure itself but places only a restriction on the $\pt$ range of the extrapolated function obtained from the scaling procedure.

\begin{figure*}
  \begin{minipage}{\textwidth}
      \includegraphics[width=0.95\textwidth]{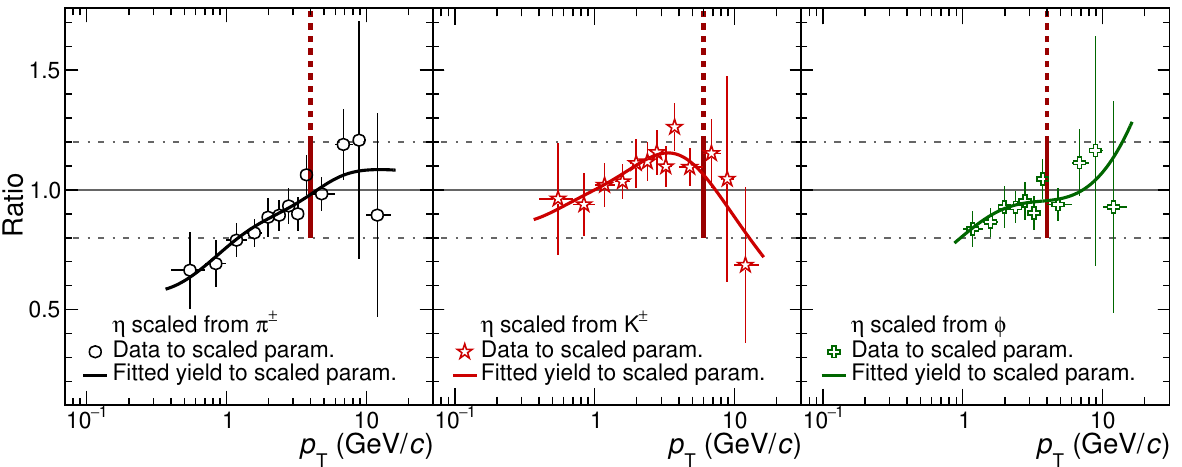}
      \caption{Ratio of the measured $\eta$ yield~(points) and its fit (lines) to the scaled parametrization obtained from different reference particles.
               The $\pt$ thresholds for the obtained scaling constants ($C$) are shown as vertical red line and reflect the point at which the ratio between scaled and reference particles is approximately flat.} 
      \label{fig:etaYieldComp}
      \vspace{0.3cm}
      \includegraphics[width=0.95\textwidth]{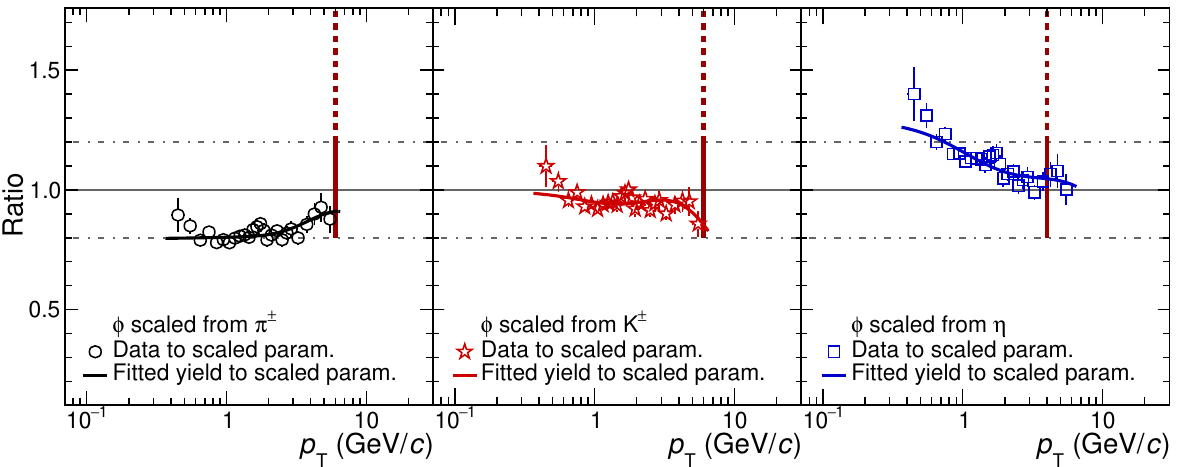}
      \caption{Ratio of the measured $\phi$ yield~(points) and its fit (lines) to the scaled parametrization obtained from different reference particles.
               The $\pt$ thresholds for the obtained scaling constants ($C$) are shown as vertical red line and reflect the point at which the ratio between scaled and reference particles is approximately flat.}
      \label{fig:phiYieldComp}
      \vspace{0.3cm}
      \includegraphics[width=0.95\textwidth]{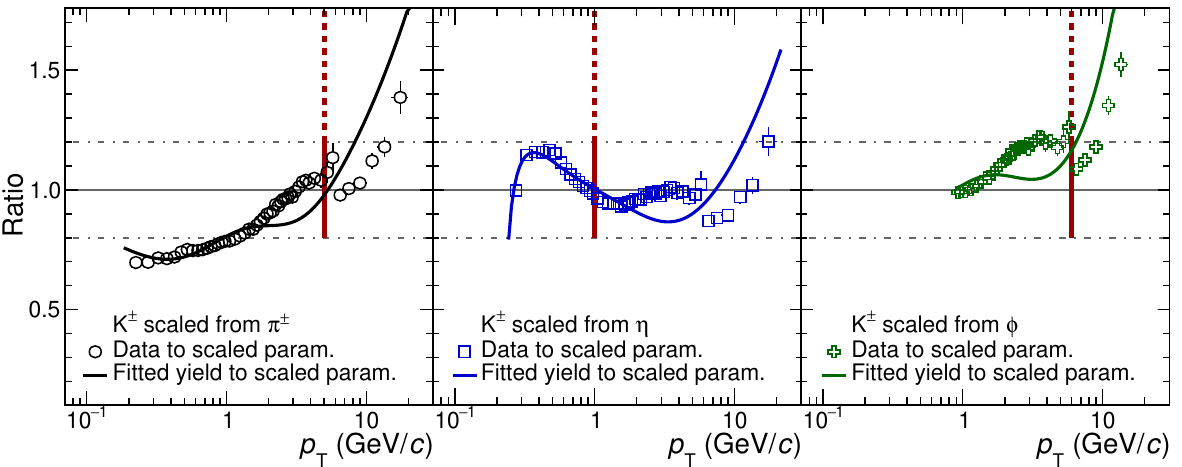}
      \caption{Ratio of the measured charged kaon yield~(points) and its fit (lines) to the scaled parametrization obtained from different reference particles.
               The $\pt$ thresholds for the obtained scaling constants ($C$) are shown as vertical red line and reflect the point at which the ratio between scaled and reference particles is approximately flat.} 
      \label{fig:chargedKaonYieldComp}
  \end{minipage}
\end{figure*}

Figures~\ref{fig:etaYieldComp} to~\ref{fig:chargedKaonYieldComp} show the comparison of the different parametrizations obtained from the $\mt$-scaling approach formulated in \Eq{eq:scalingRelation} to the parametrizations of the corresponding particle yields. 
The performance of the scaling is quantified as the ratio between the measured yield~(shown as points) and the scaled parametrization obtained from different reference particles, and additionally with the ratio of the fit (shown as lines) to the scaled parametrization.
The red vertical lines mark the different $\pt$ thresholds, which were used to establish the scaling constant and which depend on the masses of the particles, that were involved in the scaling and the $\pT$ reach of the data, while the horizontal lines mark $\pm20$\% deviation of the ratio from unity.
Below the threshold\co{, i.e.\ in the region in which the scaled spectra would extrapolate the data}, we find that the ratio generally differs significantly from the measured spectra, and hence one can conclude that the scaling does not work for low transverse momenta, i.e.\ in the region significantly lower than the threshold used to determine the scaling.
In particular, the $\eta$ meson and charged kaon can not be well reproduced by scaling the charged pions, which is also visible in \Figsm{fig:ratios1}{fig:ratioToNPi}.\\
\indent If a heavier reference particle is used, on the other hand, like the charged kaon or the $\eta$ meson the scaling performs better and the spectra can be reproduced down to lower transverse momenta. 
This is in particular the case when scaling the $\eta$ meson from the charged kaon and vice versa due to their similar mass.
It can also be observed for the $\phi$ meson, if it is scaled from the charged kaon or $\eta$ meson, respectively.
This supports the assumption that the feed-down from heavier resonances explicitly breaks the scaling from the pions and that this has a smaller influence when using the charged kaons or eta mesons as reference particle.
However, it has to be ensured that the high $\pt$ scaling factor $C$ can be obtained with sufficient precision, which was not possible using only the data measured in \pp\ collisions at $\sqrt{s}=7$~TeV.
Similar conclusions can be drawn from the comparisons of the calculated and measured particle ratios in \Figsm{fig:ratios1}{fig:ratioToNPi} in case the desired particle has been obtained from the parametrization of the charged kaon (red) or $\eta$ meson (blue) yields.
If the charged kaon and $\eta$ meson were obtained from the $\phi$ meson parameterization (green) the description of the data is worse than for other combinations, which can be explained by the imprecise $C$ values as well as the limited transverse momentum reach of the respective yield measurement of the $\phi$ meson. \\
\indent Additionally, it can be seen that for parameterizations with a similar $n$ of the power-law the ratio flattens at higher momenta, while a residual slope can be observed if different $n$ values are obtained for the parameterizations of the particles divided in the ratio. 
This implies that the scaling relation can only hold for particle spectra, which have a similar power-law behavior at high transverse momenta, as predicted.
It is reflected in the agreement between the scaled curves and the actual parameterizations above the chosen $\pt$ thresholds for those particles, which have a similar $n$ according to \Tab{tab:particlesAndParameters}. 
In these cases the agreement between the data and the parameterizations is usually better than $5-10$\%.
For the charged kaons, on the hand, a strong $\pt$-dependent difference at high transverse momenta is observed.
We have verified that the scaling at high $\pt$ cannot be restored fully by removing the step observed in \Fig{fig:ratiosYieldsToFits} through shifting the low and high $\pt$ points down and up, respectively. 
For \pp\ collisions at $\sqrt{s}=2.76$~TeV, however, the ratio of K$^\pm/\pi^\pm$ \cite{Adam:2016dau} appears to be constant above $4$~GeV/$c$ and approaches a similar value as obtained for \pp\ collisions at $\sqrt{s}=7$~TeV, which is shown in \Fig{fig:ratios1}.
Hence, it is likely that the apparent deviation of the $\text{K}^{\pm}/\pi^{\pm}$ from $\approx 0.5$ is just driven by the two highest $\pt$ points, but with the data one can also not rule out that it is broken at high $\pt$ in \pp\ collisions at 7 TeV.

\section{Summary}
\label{sec:summary}
We have presented a study on the applicability of transverse mass scaling for identified particle spectra in proton-proton collisions at $\sqrts=7$~TeV based on data taken by the ALICE experiment at the LHC. 
The measured yields of $\pi^{\pm}$, $\eta$, $\phi$ and K$^{\pm}$ mesons were parametrized and compared to estimates obtained from a generalized transverse mass scaling approach applied to different reference particle spectra.
Generalized transverse mass scaling is found not to be able to describe the measured spectra over the full range in transverse momentum, especially not when pions are used as reference particles. 
At low $\pt$, deviations of $20$\% or more are observed.
A better scaling performance is obtained, when kaons are used as reference particles.
At high $\pt$ all spectra with the possible exception of the charged kaons exhibit a scaling behavior.
Investigating the feed-down contributions from resonance decays to the charged pion yields reveals that a general scaling may not be expected when using them as reference. 
A similar study should be done for heavier reference particles, in particular for kaons, which within the available statistics exhibit a potential to be used as reference particle instead of pions\co{ for the scaling}.
Our findings imply that for precision measurements of direct photon and di-electron spectra at low transverse momentum one should measure the relevant hadronic background, instead of relying on $\mt$ scaling for its estimate.

\section*{Acknowledgments}
\label{sec:acknowledgments}
The work of L.A.\ is supported by the Research Council of Norway, Norway.
The works of C.L.\ and N.S.\ are supported by the U.S. Department of Energy, Office of Science, Office of Nuclear Physics, under contract number DE-AC05-00OR22725.

\cleardoublepage

\bibliography{biblio} 

\providecommand{\href}[2]{#2}\begingroup\raggedright\begin{thebibliography}{10}

\bibitem{Hagedorn:1965st}
R.~Hagedorn, ``{Statistical thermodynamics of strong interactions at
  high-energies},''
{\em Nuovo Cim. Suppl.} {\bfseries 3} (1965) 147--186.

\bibitem{Deutschmann:1974ne}
{\bfseries Aachen-Berlin-Bonn-CERN-Cracow-Heidelberg-London-Vienna-Warsaw}
  Collaboration, M.~Deutschmann {\em et~al.}, ``{Transverse spectra in
  $\pi^{\pm}$p and $K^{-}$p interactions between $8$~GeV/$c$ and $16$
  GeV/$c$},''
\href{http://dx.doi.org/10.1016/0550-3213(74)90473-8}{{\em Nucl. Phys.}
  {\bfseries B70} (1974) 189--204}.

\bibitem{Bartke:1976zj}
{\bfseries Aachen-Berlin-Bonn-CERN-Cracow-Heidelberg-Warsaw} Collaboration,
  J.~Bartke {\em et~al.}, ``{Simplicity of transverse energy spectra of
  hadrons},''
\href{http://dx.doi.org/10.1016/0550-3213(77)90092-X}{{\em Nucl. Phys.}
  {\bfseries B120} (1977) 14--22}.

\bibitem{Gatoff:1992cv}
G.~Gatoff and C.~Y. Wong, ``{Origin of the soft $\pt$ spectra},''
\href{http://dx.doi.org/10.1103/PhysRevD.46.997}{{\em Phys. Rev.} {\bfseries
  D46} (1992) 997--1006}.

\bibitem{Alper:1974rw}
{\bfseries British-Scandinavian} Collaboration, B.~Alper {\em et~al.}, ``{The
  production of charged Particles with high transverse momentum in
  proton-proton collisions at the CERN ISR},''
\href{http://dx.doi.org/10.1016/0550-3213(75)90248-5}{{\em Nucl. Phys.}
  {\bfseries B87} (1975) 19}.

\bibitem{Bourquin:1976fe}
M.~Bourquin and J.~M. Gaillard, ``{A simple phenomenological description of
  hadron production},''
\href{http://dx.doi.org/10.1016/0550-3213(76)90592-7}{{\em Nucl. Phys.}
  {\bfseries B114} (1976) 334--364}.

\bibitem{Albrecht:1995ug}
{\bfseries WA80} Collaboration, R.~Albrecht {\em et~al.}, ``{Production of
  $\eta$ mesons in $200$ A/GeV \SuSu\ and \SAu\ reactions},''
  \href{http://dx.doi.org/10.1016/0370-2693(95)01166-N}{{\em Phys. Lett.}
  {\bfseries B361} (1995) 14--20},
\href{http://arxiv.org/abs/hep-ex/9507009}{{\ttfamily arXiv:hep-ex/9507009
  [hep-ex]}}.

\bibitem{Abelev:2006cs}
{\bfseries STAR} Collaboration, B.~I. Abelev {\em et~al.}, ``{Strange particle
  production in \pp\ collisions at $\sqrt{s}=200$ GeV},''
  \href{http://dx.doi.org/10.1103/PhysRevC.75.064901}{{\em Phys. Rev.}
  {\bfseries C75} (2007) 064901},
\href{http://arxiv.org/abs/nucl-ex/0607033}{{\ttfamily arXiv:nucl-ex/0607033
  [nucl-ex]}}.

\bibitem{Adare:2010fe}
{\bfseries PHENIX} Collaboration, A.~Adare {\em et~al.}, ``{Measurement of
  neutral mesons in \pp\ collisions at $\sqrt(s)$= 200 GeV and scaling
  properties of hadron production},''
  \href{http://dx.doi.org/10.1103/PhysRevD.83.052004}{{\em Phys. Rev.}
  {\bfseries D83} (2011) 052004},
\href{http://arxiv.org/abs/1005.3674}{{\ttfamily arXiv:1005.3674 [hep-ex]}}.

\bibitem{Khandai:2012xx}
P.~K. Khandai, P.~Sett, P.~Shukla, and V.~Singh, ``{Transverse mass spectra and
  scaling of hadrons at RHIC and LHC energies},''
\href{http://arxiv.org/abs/1205.0648}{{\ttfamily arXiv:1205.0648 [hep-ph]}}.

\bibitem{SchaffnerBielich:2001qj}
J.~Schaffner-Bielich, D.~Kharzeev, L.~D. McLerran, and R.~Venugopalan,
  ``{Generalized scaling of the transverse mass spectrum at the relativistic
  heavy ion collider},''
  \href{http://dx.doi.org/10.1016/S0375-9474(02)00677-2}{{\em Nucl. Phys.}
  {\bfseries A705} (2002) 494--507},
\href{http://arxiv.org/abs/nucl-th/0108048}{{\ttfamily arXiv:nucl-th/0108048
  [nucl-th]}}.

\bibitem{Loizides:2016tew}
C.~Loizides, ``{Experimental overview on small collision systems at the LHC},''
  \href{http://dx.doi.org/10.1016/j.nuclphysa.2016.04.022}{{\em Nucl. Phys.}
  {\bfseries A956} (2016) 200--207},
\href{http://arxiv.org/abs/1602.09138}{{\ttfamily arXiv:1602.09138 [nucl-ex]}}.

\bibitem{Adare:2012yt}
{\bfseries PHENIX} Collaboration, A.~Adare {\em et~al.}, ``{Direct photon
  production in \pp\ collisions at $\sqrt{s}=200$ GeV at midrapidity},''
  \href{http://dx.doi.org/10.1103/PhysRevD.86.072008}{{\em Phys. Rev.}
  {\bfseries D86} (2012) 072008},
\href{http://arxiv.org/abs/1205.5533}{{\ttfamily arXiv:1205.5533 [hep-ex]}}.

\bibitem{Adare:2012vn}
{\bfseries PHENIX} Collaboration, A.~Adare {\em et~al.}, ``{Direct photon
  production in \dAu collisions at $\snn=200$ GeV},''
  \href{http://dx.doi.org/10.1103/PhysRevC.87.054907}{{\em Phys. Rev.}
  {\bfseries C87} (2013) 054907},
\href{http://arxiv.org/abs/1208.1234}{{\ttfamily arXiv:1208.1234 [nucl-ex]}}.

\bibitem{Afanasiev:2012dg}
{\bfseries PHENIX} Collaboration, S.~Afanasiev {\em et~al.}, ``{Measurement of
  direct photons in \AuAu\ collisions at $\snn=200$ GeV},''
  \href{http://dx.doi.org/10.1103/PhysRevLett.109.152302}{{\em Phys. Rev.
  Lett.} {\bfseries 109} (2012) 152302},
\href{http://arxiv.org/abs/1205.5759}{{\ttfamily arXiv:1205.5759 [nucl-ex]}}.

\bibitem{Adam:2015lda}
{\bfseries ALICE} Collaboration, J.~Adam {\em et~al.}, ``{Direct photon
  production in \PbPb\ collisions at $\snn=2.76$ TeV},''
  \href{http://dx.doi.org/10.1016/j.physletb.2016.01.020}{{\em Phys. Lett.}
  {\bfseries B754} (2016) 235--248},
\href{http://arxiv.org/abs/1509.07324}{{\ttfamily arXiv:1509.07324 [nucl-ex]}}.

\bibitem{Adare:2009qk}
{\bfseries PHENIX} Collaboration, A.~Adare {\em et~al.}, ``{Detailed
  measurement of the $e^+ e^-$ pair continuum in \pp\ and \AuAu\ collisions at
  $\snn=200$ GeV and implications for direct photon production},''
  \href{http://dx.doi.org/10.1103/PhysRevC.81.034911}{{\em Phys. Rev.}
  {\bfseries C81} (2010) 034911},
\href{http://arxiv.org/abs/0912.0244}{{\ttfamily arXiv:0912.0244 [nucl-ex]}}.

\bibitem{Adamczyk:2015lme}
{\bfseries STAR} Collaboration, L.~Adamczyk {\em et~al.}, ``{Measurements of
  di-electron production in \AuAu\ collisions at $\snn=200$ GeV from the STAR
  experiment},'' \href{http://dx.doi.org/10.1103/PhysRevC.92.024912}{{\em Phys.
  Rev.} {\bfseries C92} no.~2, (2015) 024912},
\href{http://arxiv.org/abs/1504.01317}{{\ttfamily arXiv:1504.01317 [hep-ex]}}.

\bibitem{Adam:2016dau}
{\bfseries ALICE} Collaboration, J.~Adam {\em et~al.}, ``{Multiplicity
  dependence of charged pion, kaon, and (anti)proton production at large
  transverse momentum in p-Pb collisions at $\snn=5.02$ TeV},''
  \href{http://dx.doi.org/10.1016/j.physletb.2016.07.050}{{\em Phys. Lett.}
  {\bfseries B760} (2016) 720--735},
\href{http://arxiv.org/abs/1601.03658}{{\ttfamily arXiv:1601.03658 [nucl-ex]}}.

\bibitem{Abelev:2012cn}
{\bfseries ALICE} Collaboration, B.~Abelev {\em et~al.}, ``{Neutral pion and
  $\eta$ meson production in proton-proton collisions at $\sqrt{s}=0.9$ TeV and
  $\sqrt{s}=7$ TeV},''
  \href{http://dx.doi.org/10.1016/j.physletb.2012.09.015}{{\em Phys. Lett.}
  {\bfseries B717} (2012) 162--172},
\href{http://arxiv.org/abs/1205.5724}{{\ttfamily arXiv:1205.5724 [hep-ex]}}.

\bibitem{Abelev:2012hy}
{\bfseries ALICE} Collaboration, B.~Abelev {\em et~al.}, ``{Production of
  $K^*(892)^0$ and $\phi(1020)$ in \pp\ collisions at $\sqrt{s}=7$ TeV},''
  \href{http://dx.doi.org/10.1140/epjc/s10052-012-2183-y}{{\em Eur. Phys. J.}
  {\bfseries C72} (2012) 2183},
\href{http://arxiv.org/abs/1208.5717}{{\ttfamily arXiv:1208.5717 [hep-ex]}}.

\bibitem{Abelev:2012jp}
{\bfseries ALICE} Collaboration, B.~Abelev {\em et~al.}, ``{Multi-strange
  baryon production in \pp\ collisions at $\sqrt{s} = 7$ TeV with ALICE},''
  \href{http://dx.doi.org/10.1016/j.physletb.2012.05.011}{{\em Phys. Lett.}
  {\bfseries B712} (2012) 309--318},
\href{http://arxiv.org/abs/1204.0282}{{\ttfamily arXiv:1204.0282 [nucl-ex]}}.

\bibitem{Abelev:2014qqa}
{\bfseries ALICE} Collaboration, B.~B. Abelev {\em et~al.}, ``{Production of
  $\Sigma(1385)^{\pm}$ and $\Xi(1530)^{0}$ in proton-proton collisions at
  $\sqrt{s}=$ 7 TeV},''
  \href{http://dx.doi.org/10.1140/epjc/s10052-014-3191-x}{{\em Eur. Phys. J.}
  {\bfseries C75} no.~1, (2015) 1},
\href{http://arxiv.org/abs/1406.3206}{{\ttfamily arXiv:1406.3206 [nucl-ex]}}.

\bibitem{Peresunko:2012tt}
{\bfseries ALICE} Collaboration, D.~Peresunko, ``{Neutral meson production in
  pp and Pb-Pb collisions at the LHC measured with ALICE},''
  \href{http://dx.doi.org/10.1016/j.nuclphysa.2013.02.127}{{\em Nucl. Phys.}
  {\bfseries A904-905} (2013) 755c--758c},
\href{http://arxiv.org/abs/1210.5749}{{\ttfamily arXiv:1210.5749 [nucl-ex]}}.

\bibitem{Riabov:2017jig}
{\bfseries ALICE} Collaboration, V.~G. Riabov, ``{Resonance production in
  ALICE},''
\href{http://dx.doi.org/10.1088/1742-6596/798/1/012054}{{\em J. Phys. Conf.
  Ser.} {\bfseries 798} no.~1, (2017) 012054}.

\bibitem{Cleymans:2012ya}
J.~Cleymans and D.~Worku, ``{Relativistic Thermodynamics: Transverse momentum
  distributions in high-energy physics},''
  \href{http://dx.doi.org/10.1140/epja/i2012-12160-0}{{\em Eur. Phys. J.}
  {\bfseries A48} (2012) 160},
\href{http://arxiv.org/abs/1203.4343}{{\ttfamily arXiv:1203.4343 [hep-ph]}}.

\bibitem{Klaus}
K.~Reygers, ``private communication,''  (2016) .

\bibitem{Schnedermann:1993ws}
E.~Schnedermann, J.~Sollfrank, and U.~W. Heinz, ``{Thermal phenomenology of
  hadrons from 200 A/GeV \SS\ collisions},''
  \href{http://dx.doi.org/10.1103/PhysRevC.48.2462}{{\em Phys. Rev.} {\bfseries
  C48} (1993) 2462--2475},
\href{http://arxiv.org/abs/nucl-th/9307020}{{\ttfamily arXiv:nucl-th/9307020
  [nucl-th]}}.

\bibitem{Acharya:2017tlv}
{\bfseries ALICE} Collaboration, S.~Acharya {\em et~al.}, ``{$\pi^0$ and $\eta$
  meson production in proton-proton collisions at $\sqrt{s}=8$ TeV},''
\href{http://arxiv.org/abs/1708.08745}{{\ttfamily arXiv:1708.08745 [hep-ex]}}.

\bibitem{Abelev:2014laa}
{\bfseries ALICE} Collaboration, B.~B. Abelev {\em et~al.}, ``{Production of
  charged pions, kaons and protons at large transverse momenta in \pp\ and
  \PbPb\ collisions at $\snn = 2.76$ TeV},''
  \href{http://dx.doi.org/10.1016/j.physletb.2014.07.011}{{\em Phys. Lett.}
  {\bfseries B736} (2014) 196--207},
\href{http://arxiv.org/abs/1401.1250}{{\ttfamily arXiv:1401.1250 [nucl-ex]}}.

\bibitem{Acharya:2017hyu}
{\bfseries ALICE} Collaboration, S.~Acharya {\em et~al.}, ``{Production of
  ${\pi ^0}$ and $\eta $ mesons up to high transverse momentum in pp collisions
  at 2.76 TeV},'' \href{http://dx.doi.org/10.1140/epjc/s10052-017-4890-x}{{\em
  Eur. Phys. J.} {\bfseries C77} no.~5, (2017) 339},
\href{http://arxiv.org/abs/1702.00917}{{\ttfamily arXiv:1702.00917 [hep-ex]}}.

\bibitem{Sjostrand:2014zea}
T.~Sjöstrand, S.~Ask, J.~R. Christiansen, R.~Corke, N.~Desai, P.~Ilten,
  S.~Mrenna, S.~Prestel, C.~O. Rasmussen, and P.~Z. Skands, ``{An Introduction
  to PYTHIA 8.2},'' \href{http://dx.doi.org/10.1016/j.cpc.2015.01.024}{{\em
  Comput. Phys. Commun.} {\bfseries 191} (2015) 159--177},
\href{http://arxiv.org/abs/1410.3012}{{\ttfamily arXiv:1410.3012 [hep-ph]}}.

\bibitem{ALICE-PUBLIC-2017-005}
{\bfseries ALICE} Collaboration, ``{The ALICE definition of primary
  particles},'' {\em ALICE-PUBLIC-2017-005} (Jun, 2017) .
  \url{https://cds.cern.ch/record/2270008}.

\bibitem{Sjostrand:2006za}
T.~Sjostrand, S.~Mrenna, and P.~Z. Skands, ``{PYTHIA 6.4 Physics and Manual},''
  \href{http://dx.doi.org/10.1088/1126-6708/2006/05/026}{{\em JHEP} {\bfseries
  05} (2006) 026},
\href{http://arxiv.org/abs/hep-ph/0603175}{{\ttfamily arXiv:hep-ph/0603175
  [hep-ph]}}.

\bibitem{Adam:2017zbf}
{\bfseries ALICE} Collaboration, J.~Adam {\em et~al.}, ``{K$^{*}(892)^{0}$ and
  $\phi(1020)$ meson production at high transverse momentum in pp and Pb-Pb
  collisions at $\sqrt{s_\mathrm{NN}}$ = 2.76 TeV},''
  \href{http://dx.doi.org/10.1103/PhysRevC.95.064606}{{\em Phys. Rev.}
  {\bfseries C95} no.~6, (2017) 064606},
\href{http://arxiv.org/abs/1702.00555}{{\ttfamily arXiv:1702.00555 [nucl-ex]}}.

\end{thebibliography}\endgroup
\bibliographystyle{utphys}
\end{document}